\documentclass[10pt,conference]{IEEEtran}
\usepackage{graphicx, graphics,epsfig}
\usepackage{color}
\usepackage[cmex10]{amsmath}
\usepackage{amssymb}
\usepackage{url}
\usepackage{amsfonts}
\usepackage{latexsym}
\usepackage{tabularx}
\usepackage{stmaryrd}
\usepackage{bbm}

\usepackage{subfigure,dsfont,comment,colortbl,cite}
\usepackage{algorithm}
\usepackage{algorithmic}

\usepackage{txfonts}

\ifodd 1
\newcommand{\com}[1]{\textbf{\color{blue} (COMMENT: #1)}} 
\else

\newcommand{\com}[1]{}
\fi

\newcommand{\beq}   {\begin{equation}}
\newcommand{\eeq}   {\end{equation}}
\newcommand{\bea}   {\begin{eqnarray}}
\newcommand{\eea}   {\end{eqnarray}}
\newcommand{\bda}   {\begin{eqnarray*}}
\newcommand{\eda}   {\end{eqnarray*}}
\newcommand{\bdalign}   {\begin{align*}}
\newcommand{\edalign}   {\end{align*}}

\newtheorem{theorem}{\bf Theorem}

\newtheorem{proposition}{\bf Proposition}

\title{
Optimal Distributed P2P Streaming under Node Degree Bounds
}

\author{{\large Shaoquan Zhang$^{\ast}$, Ziyu Shao$^{\ast}$, Minghua Chen$^{\ast}$, and Libin Jiang$^{\dagger}$}
\\
$^{\ast}$Department of Information Engineering, The Chinese University of
Hong Kong
\\
$^{\dagger}$Department of EECS, University of California, Berkeley
\\
Email: $^{\ast}$\{zsq008, zyshao6, minghua\}@ie.cuhk.edu.hk,$^{\dagger}$\{libinj\}@caltech.edu }

\begin{document}
\maketitle

\begin{abstract}
We study the problem of maximizing the broadcast rate in peer-to-peer
(P2P) systems under \emph{node degree bounds}, i.e., the number of
neighbors a node can simultaneously connect to is upper-bounded. The
problem is critical for supporting high-quality video streaming in
P2P systems, and is challenging due to its combinatorial nature. In
this paper, we address this problem by providing the first distributed
solution that achieves near-optimal broadcast rate under arbitrary
node degree bounds, and over arbitrary overlay graph. It runs on individual
nodes and utilizes only the measurement from their one-hop neighbors,
making the solution easy to implement and adaptable to peer churn
and network dynamics. Our solution consists of two distributed algorithms
proposed in this paper that can be of independent interests: a network-coding
based broadcasting algorithm that optimizes the broadcast rate given
a topology, and a Markov-chain guided topology hopping algorithm that optimizes
the topology. Our distributed broadcasting algorithm achieves the
optimal broadcast rate over arbitrary P2P topology, while previously
proposed distributed algorithms obtain optimality only for P2P complete
graphs. We prove the optimality of our solution and its convergence
to a neighborhood around the optimal equilibrium under noisy measurements
or without time-scale separation assumptions. We demonstrate the effectiveness
of our solution in simulations using uplink bandwidth statistics of
Internet hosts.
\end{abstract}

\begin{table*}
\caption{Summary and comparison of previous work and this work for maximizing
P2P broadcast rate.}

\label{tab:summary} \begin{small}

\begin{centering}
\begin{tabular}{|c|c|c|c|c|}
\hline
References\emph{ }  & General\emph{ }  & Arbitrary Node\emph{ }  & Exact or\emph{ }$1-\epsilon$  & Distributed \tabularnewline
 & Overlay Graph?  & Degree Bound?  & Optimality?  & Solution?\tabularnewline
\hline
\hline
Mutualcast \cite{all:Mutualcast:LPZ05} and the algorithms in \cite{massoulie2007rdb,all:P2PStreaming:KLR.07}  & $\times$  & $\times$  & \textbf{$\mathbf{\checkmark}$}  & \textbf{$\mathbf{\checkmark}$}\tabularnewline
\hline
Iterative in \cite{yicui06_optimal,sudipta2009lcclc}  & \textbf{$\mathbf{\checkmark}$}  & $\times$  & \textbf{$\mathbf{\checkmark}$}  & $\times$\tabularnewline
\hline
CoopNet/SplitStream \cite{castro2003shb,padmanabhan2002ccn}  & $\times$  & \textbf{$\mathbf{\checkmark}$}  & $\times$  & $\times$\tabularnewline
\hline
ZIGZAG~\cite{zigzag}, PRIME~\cite{magharei2009prime}  & \textbf{$\mathbf{\checkmark}$}  & \textbf{$\mathbf{\checkmark}$}  & $\times$  & \textbf{$\mathbf{\checkmark}$}\tabularnewline
\hline
Cluster-tree \cite{streaming_capacity.icdcs10}  & $\times$  & \textbf{$\mathbf{\checkmark}$}  & conditionally optimal $^{*}$  & $\times$\tabularnewline
\hline
\textbf{This paper }  & \textbf{$\mathbf{\checkmark}$}  & \textbf{$\mathbf{\checkmark}$}  & \textbf{$\mathbf{\checkmark}$}  & \textbf{$\mathbf{\checkmark}$}\tabularnewline
\hline
\end{tabular}
\par\end{centering}

\end{small}
\begin{quote}
\begin{scriptsize} $\;{}^{*}$ The Cluster-Tree algorithm is $(1-\epsilon)$-optimal
with high probability if the node degree bound is $O\left(\log N\right)$.
\end{scriptsize}
\end{quote}

\end{table*}

\section{Introduction}

\label{sec:Intro}

Peer-to-peer (P2P) systems have provided a scalable and cost effective
way for streaming video in the past decade. Recent studies \cite{fenglili,AbeKirCig,wang2007r,zhang2005coolstreaming},
however, indicate that the practical performance of P2P streaming
systems can be far from their theoretical optimal.

There have been work studying the performance limit of P2P systems
to understand and unleash their potential. One focus is on the \emph{streaming
capacity} problem \cite{streaming_capacity.allerton09} in P2P live streaming systems
, i.e., maximizing the streaming rate subject to the peering and overlay topology constraints.
The problem is critical for supporting high-quality video, which is
determined by the streaming rate, in P2P live streaming systems. In this paper, we
focus on the broadcast scenario where all peers in the system are
receivers.

The case of unconstrained peering on top of a complete graph is well
studied, where the maximum broadcast rate is derived in several papers
\cite{all:P2PStreaming:KLR.07,all:Mutualcast:LPZ05,all:NetCodP2P:CYHF.06,massoulie2007rdb,chen2008ump}.
The case of unconstrained peering over general graph can also be addressed
by using a centralized solution\cite{sudipta2009lcclc}.

The streaming capacity problem becomes NP-Complete over general graph
with \emph{node degree bounds} \cite{streaming_capacity.icdcs10}.
Node degree is defined as the number of simultaneous active connections that a node maintains with its neighbors. Due to connection overhead costs, it is necessary
to limit the number of simultaneous connections a peer can maintain.
This naturally bounds the node degrees in P2P systems. For instance, in practical
systems such as PPLive \cite{all:pplive}, the total number of neighbors of
a node is usually bounded around 200, and the number
of active neighbors of a node is usually bounded by 10-15 \cite{streaming_capacity.allerton09}. In such large P2P systems with hundreds of thousands of peers, the system topology is not a complete graph.

There has been work studying this challenging problem of maximizing streaming rate under node degree bounds and over general P2P graph. SplitStream/CoopNet~\cite{castro2003shb,padmanabhan2002ccn},
ZIGZAG~\cite{zigzag}, PRIME~\cite{magharei2009prime} and most
practical systems (such as PPLive \cite{all:pplive} and UUSee \cite{all:uusee})
bound node degree but do not provide rate optimality guarantee. Recently,
the authors in \cite{streaming_capacity.icdcs10} proposed a centralized
Cluster-Tree algorithm that achieves near-optimal broadcast rate with
high probability over complete graph, under the assumption that the
node degree bound is at least logarithmic in the size of the network. A summary
and comparison of previous work and this work are in Table \ref{tab:summary}.

Despite of these exciting results, the following two important questions
remain open:
\begin{itemize}
\item What is the maximum broadcast rate under arbitrary node degree bounds,
and over general P2P overlay graph?
\item How to achieve the maximum broadcast rate in a \emph{distributed}
manner?
\end{itemize}
Systems running distributed algorithms, compared with those running
centralized algorithms, are more adaptable to peer churn and network
dynamics.

In this paper, we answer the above two questions and make the following
contributions:
\begin{itemize}
\item We provide the first distributed solution that achieves a broadcast
rate arbitrarily close to the optimal under arbitrary node degree
bounds, and over arbitrary overlay graph. Our solution runs on individual
nodes and utilizes only the information from their one-hop neighbors.
\end{itemize}
Our solution consists of the following two algorithms that can be
of independent interests.
\begin{itemize}
\item We propose a distributed broadcasting algorithm that achieves the
optimal broadcast rate over arbitrary overlay graph. Previous distributed
P2P broadcasting algorithms are optimal only for complete overlay
graph \cite{massoulie2007rdb,all:Mutualcast:LPZ05,all:P2PStreaming:KLR.07}.
Our algorithm is based on network coding and utilizes back-pressure
arguments.
\item We also propose a distributed algorithm that optimizes the topology.
In this algorithm, each node hops among their possible set of neighbors
towards the best peering configuration. Our algorithm is inspired
by a set of log-sum-exp approximation and Markov chain based arguments
expounded in \cite{MA:CLSC10}.
\item We prove the optimality of the overall solution. We also prove its
convergence to a neighborhood around the optimal equilibrium in the presence of
noisy measurements or without time-scale separation assumptions. We
demonstrate the effectiveness of our solution in simulations using
uplink bandwidth statistics of Internet hosts.
\end{itemize}


\section{Problem Formulation}

\label{sec:prob.formulation}

\subsection{Settings and Notations}

We model the P2P overlay network as a general directed graph $G=(V,E)$, where
\emph{$V$} denotes the set of nodes and \emph{$E$} denotes the set
of links. Each link in the graph corresponds to a TCP/UDP
connection between two nodes. Let $N_{v}$ denote the neighbor set of node $v\in V$ in the graph. Each node $v\in V$ is associated
with an upload capacity $C_{v}\ge0$. We assume there is no constraint on the downloading
rate for each node $v\in V$. This assumption can be partly justified
by the empirical observation that as residential broadband
connections with asymmetric upload and download rates become increasingly dominant, bottlenecks
typically are at the uplinks of the access networks rather than
in the middle of the Internet.

As such, P2P networks have capacity limits on the nodes instead of links. This is different from traditional underlay networks where the capacity limits are on the links.

We focus on the single-source streaming scenario, i.e., a source
$s$ broadcasts a continuous stream of contents to the entire network; we
denote its receiver set as $R\triangleq V-\{s\}$.

\begin{figure}[hbt!]
\centering
\subfigure[]{\includegraphics[width=0.17\textwidth]{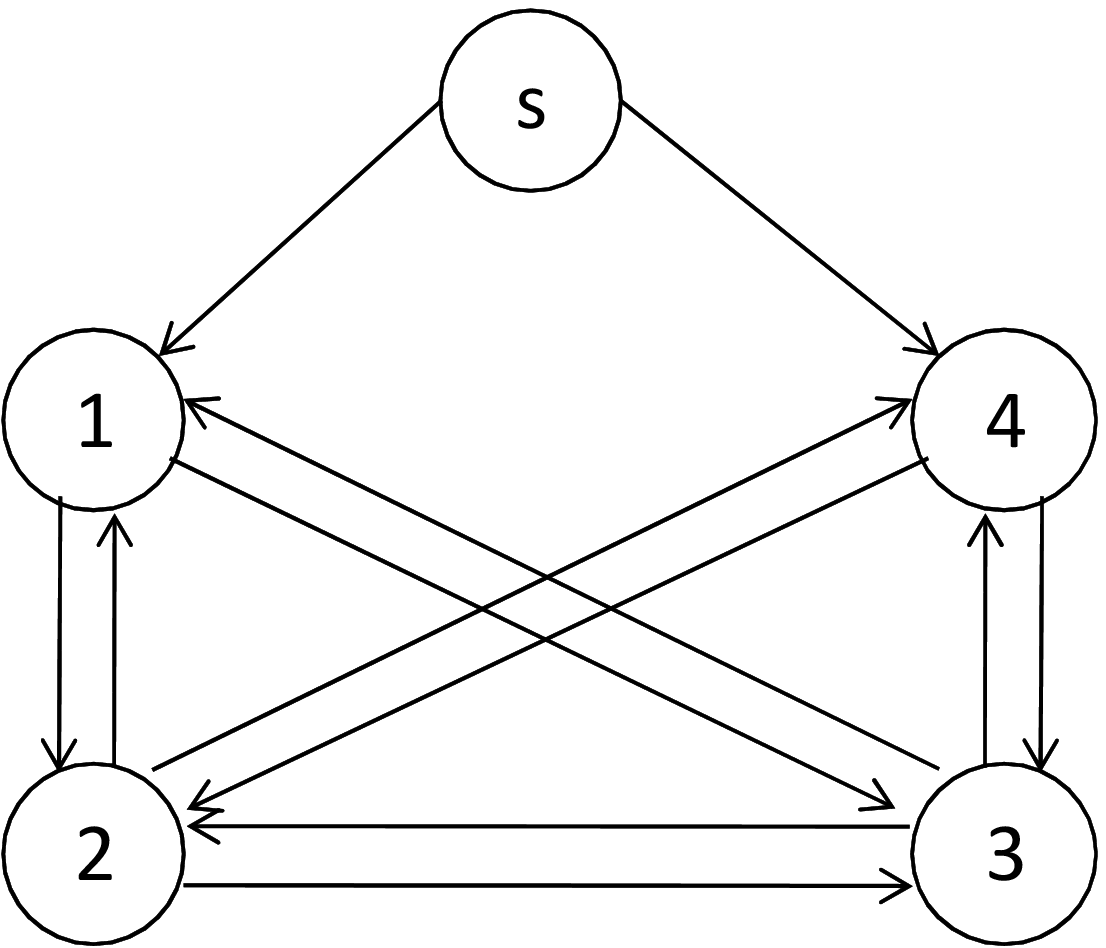}}
\subfigure[]{\includegraphics[width=0.17\textwidth]{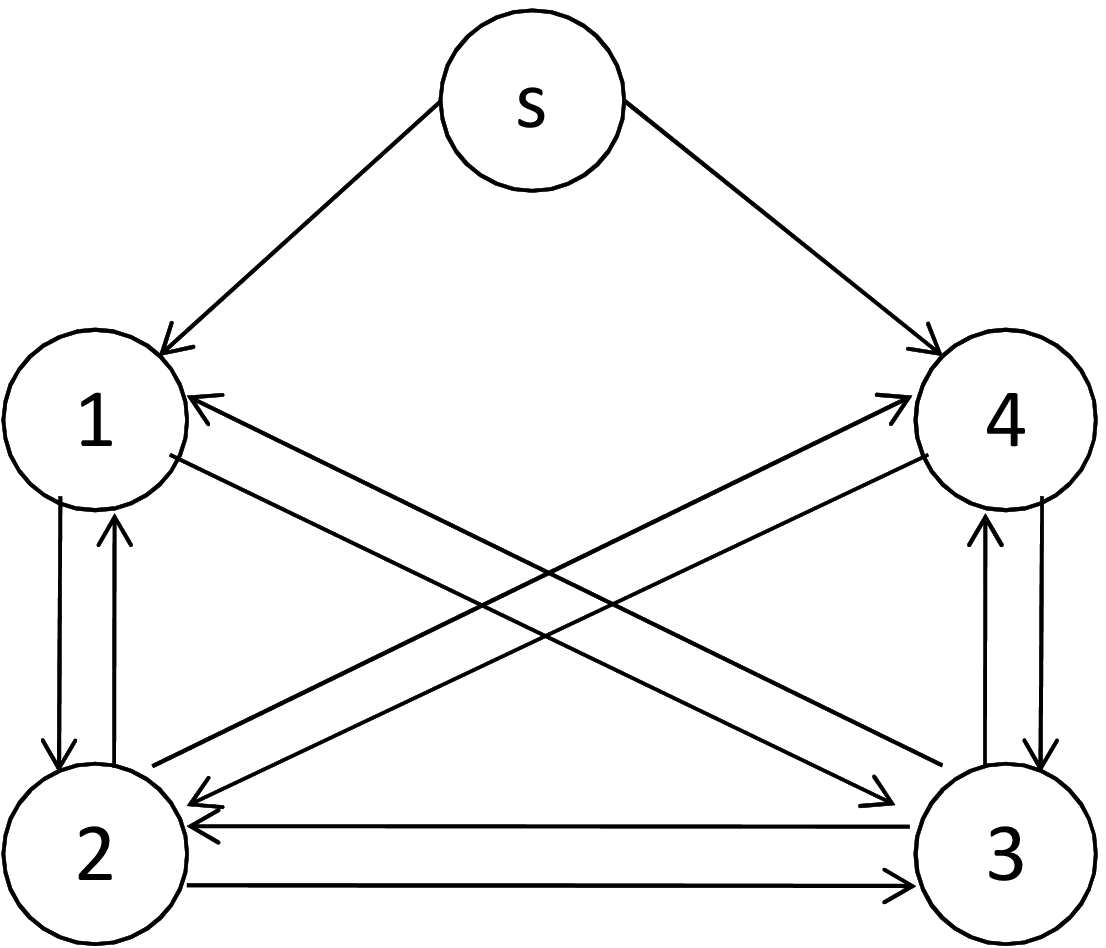}}\\
\subfigure[]{\includegraphics[width=0.17\textwidth]{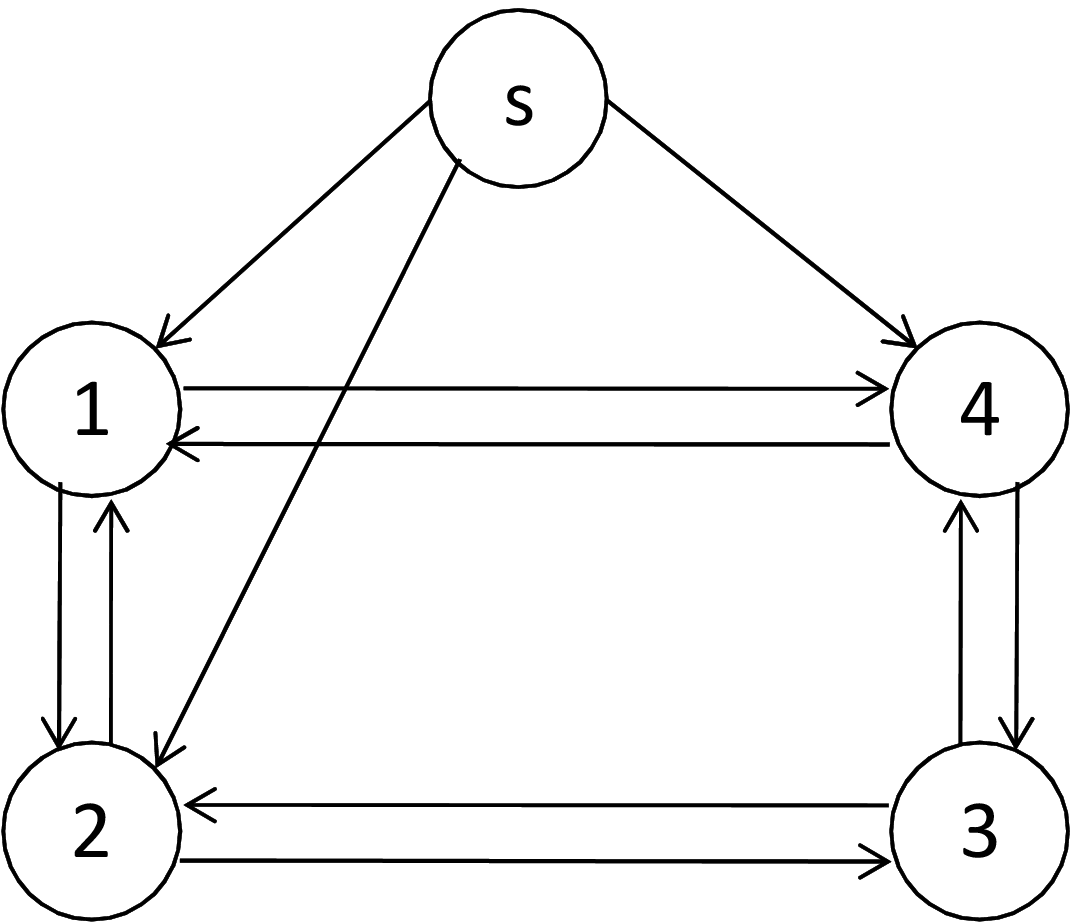}}
\subfigure[]{\includegraphics[width=0.17\textwidth]{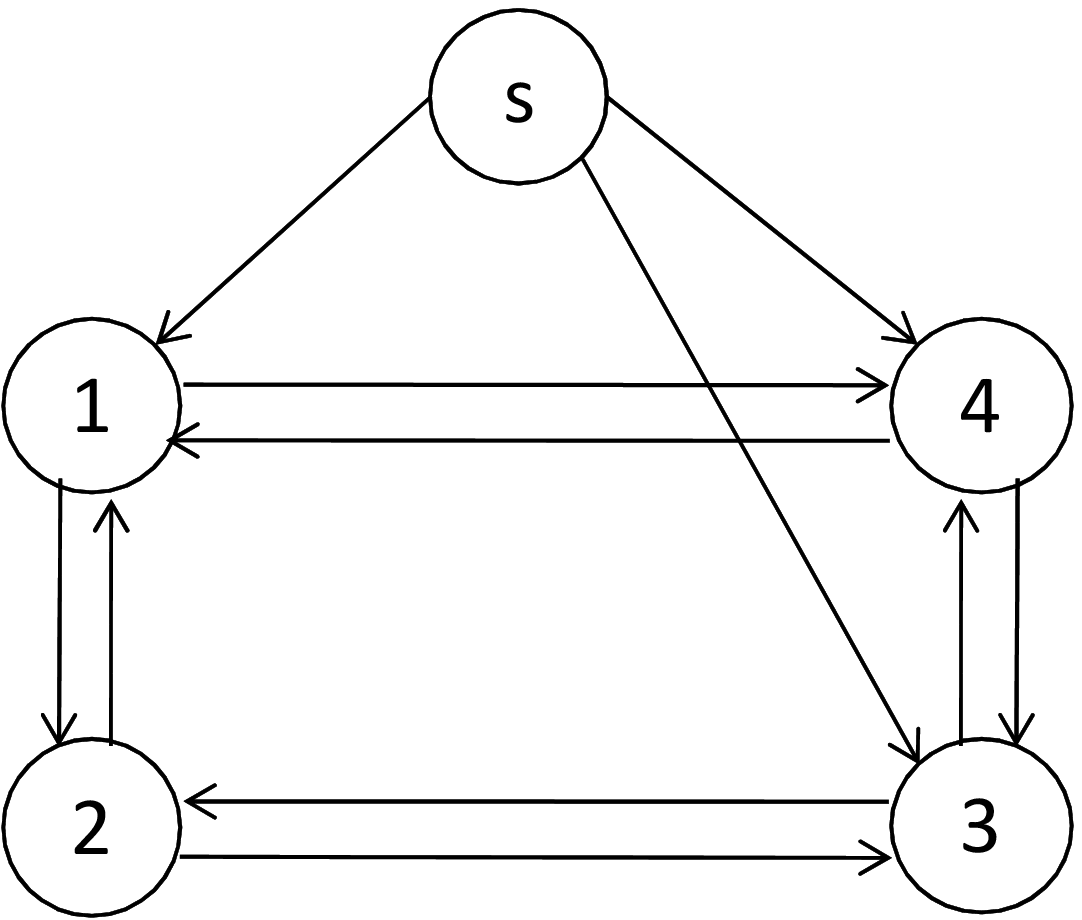}}
\caption{Peering configuration examples for a $5$-node network with node degree bound $3$ for each node.}
\label{fig:pc}
\end{figure}

We consider the peering constraints that each node has a degree bound
$B_{v}$, i.e., it can only exchange streaming content
with up to a $B_v$ number of neighbors \emph{simultaneously} due to connection overhead cost.
We allow different nodes to have different degree bounds.
Fig.~\ref{fig:pc} shows four sample peering configurations of a $5$-node
network with node degree bound $3$ for each node.

Let $\mathcal{F}$ denote the set of all feasible peering configurations
over graph $G$ under node degree bounds. Given a configuration $f\in\mathcal{F}$, we obtain a
connected sub-graph of $G$ that satisfies the node degree bound constraints. We denote this sub-graph as $G_{f}=\left(V,E_{f}\right)$, where $E_{f}$
represents the set of links in this sub-graph. We denote $N_{v,f}$
as the set of node $v$'s neighbors in this sub-graph. We have $|N_{v,f}|\leq B_{v}$
where $|\cdot|$ represents the size of a set.

\subsection{Problem Formulation and Our Approach}

For a configuration $f\in\mathcal{F}$, let $x_{f}$
be the maximum achievable broadcast rate under $f$, i.e., the highest rate at which
every node in the system can receive the streaming content simultaneously. The problem of
maximizing broadcast rate under node degree bounds can be
formulated as follows:
\begin{eqnarray}
\mathbf{MRC:} & \max_{f\in\mathcal{F}} & x_{f}.\label{MWC}
\end{eqnarray}

This problem is \emph{combinatorial} in nature which is known to be NP-complete \cite{streaming_capacity.icdcs10},
and there is no efficient approximate solution to the problem even
in a centralized manner.

In this paper, we address this problem by providing a distributed
solution. In particular, we first develop a distributed broadcasting
algorithm that can achieve $x_{f}$ under arbitrary $f\in\mathcal{F}$.
We then design a distributed algorithm that optimizes towards
the best peering configurations. They operate in tandem to achieve
a close-to-optimal broadcast rate under arbitrary node degree bounds,
and over arbitrary overlay graph. We elaborate on these two algorithms
in the following two sections.

\section{The Proposed Distributed Broadcasting Algorithm}

\label{sec:streaming_rate}


By exploiting network coding \cite{all:NetCod:ACLY00}, we design a back-pressure based distributed broadcasting algorithm. Back-pressure type algorithm is proposed initially in \cite{Tassiulas92backpressure}. This type of algorithms select a subset of queues in the system with the maximum back-pressures and serve these queues subject to resource constraints, where back-pressure is defined as the difference between the queue at the local node and that of its downstream nodes. Back-pressure algorithm design has found applications in many network resource allocation domains \cite{all:Backpressure:MAW96}, \cite{all:Backpressure:ESP05}, \cite{all:Backpressure:NMR05}. In this paper, we apply this method for the first time to design distributed P2P broadcasting algorithm. Our algorithm can achieve the maximum broadcast rate over arbitrary P2P topology.

\subsection{Routing vs. Network Coding}\label{subsec:routing-vs-coding}

In P2P systems, there are two approaches for broadcasting contents:
one is based on routing \cite{liu2008opportunities}, in which nodes only store and forward packets;
and the other is based on network coding \cite{all:NetCod:ACLY00}, \cite{liu2008opportunities},
in which a node is also allowed to mix information and output data
as functions of the data it received. Some commercial P2P systems
are built upon routing-based approach (e.g., PPLive \cite{all:pplive}),
and some are based on network coding (e.g., UUSee \cite{all:uusee,liu-uusee})%
\footnote{We refer interested readers to \cite{liu-uusee,hei2007msl}
for more details on performance of routing-based and network-coding-based
practical P2P systems. We focus on optimal distributed P2P broadcasting
algorithm design based on network coding in this paper. %
}. It is known that both routing and network coding approaches can
achieve optimal broadcast rate over arbitrary P2P graph \cite{chen2008ump,massoulie2007rdb}.
Compared to routing-based approach, the network-coding based approach introduces additional packet
header overhead for carrying coding coefficients (e.g., $3\%$ extra overhead according to \cite{chou2003practical}) and computation complexity for encoding and decoding (e.g., \cite{wang2007r,liu-uusee} discuss how to keep the complexity low). However, the network-coding based approach is robust to peer dynamics since there is no need
for constructing and maintaining the spanning trees. In this
section, we design a distributed broadcasting algorithm based on network
coding that is robust to dynamics. In Section \ref{sec:conclusion}, we will discuss how the overall problem can be solved by using centralized solutions when only routing is allowed.

\subsection{Network Coding Based Formulation}\label{subsec:nc-formulation}

According to the Max-Flow-Min-Cut theorem, a data transmission of rate
$z$ between source $s$ and a receiver $d$ is \emph{feasible}
if and only if there exists a flow, denoted as $\boldsymbol{f}^{d}$,
satisfying the following flow conservation constraints: \begin{eqnarray}
\sum_{u\in in(v)}f_{uv}^{d} & \leq & \sum_{u\in out(v)}f_{vu}^{d},\;\;\forall v\in R-\left\{ d\right\} ,\label{eq:flow-constraint-2}\\
z & \leq & \sum_{u\in out(s)}f_{su}^{d},\label{eq:eq:flow-constraint-3}\\
0 & \leq & \boldsymbol{f}^{d},\label{eq:flow-constraint-1}\end{eqnarray}
 where $in(v)\triangleq\{u|(u,v)\in E_{f}\}$ is the set of nodes
sending content to $v$ under configuration $f$, and $out(v)\triangleq\{u|(v,u)\in E_{f},u\neq s\}$
is the set of nodes receiving content from  $v$.

A powerful theorem established in \cite{all:NetCod:ACLY00} states
that a multicast or broadcast rate $z$ from
$s$ to a set of receivers is achievable if and only if $z$ is feasible
for $s$ and any receiver $d$. This is
a strong result as it says that if the network can support a unicast
rate of $z$ between $s$ and any receiver assuming other receivers'
traffic is absent, then it can support a multicast rate of $z$ to
all the receivers simultaneously. Such rate $z$ can be achieved by every
node in the network performing network coding~\cite{all:NetCod:ACLY00}. Further, authors in~\cite{ho2009dynamic,chou2003practical}
show that it is sufficient to perform random linear network coding.

In random linear network coding, by independently and randomly choosing
a set of coding coefficients from a finite field, each node sends
out the coded packet as a linear combination of the
node's received packets. The combination information is
specified by a \emph{coefficient vector} in the packet header, which is
updated by applying the same linear transformations as to the data. When
one node receives a full set of linearly independent coded packets,
it can decode and recover the original packets. In this paper, we
focus on the distributed algorithm design. The discussions of
decoding probability and implementation details
can be found in~\cite{ho2009dynamic,chou2003practical}.

Under the setting of network coding, we can consider $\boldsymbol{f}^{d}$ as a {}``virtual''
information flow between $s$ and $d$. Multiple information flows
{}``piggyback'' together to transmit over the physical links. The
actual physical rate over a physical link is only the maximum rate
of individual information flows passing over it. Let $g_{uv}$ be the physical
flow rate over a link $(u,v)\in E_{f}$, then we have $f_{uv}^{d}\leq g_{uv}$
for all $d\in R$.

With the above understanding, we formulate the problem of maximizing
broadcast rate under configuration $f$ as follows: \begin{eqnarray}
 & \mathbf{MP:} & \underset{z,\boldsymbol{f,g}\ge0}{\max}U(z)\label{eq:MP}\\
 & \mbox{s.t.} & \!\!\!\sum_{u\in in(v)}f_{uv}^{d}+z\mathbbm{1}_{\left\{ v=s\right\} }\leq\sum_{u\in out(v)}\!\!\! f_{vu}^{d},\forall v\in V-\{d\},d\in R,\label{eq:MPC1}\\
 &  & f_{vu}^{d}\leq g_{vu},\forall v\in V,\forall u\in out(v),d\in R,\label{eq:MPC2}\\
 &  & \!\!\!\sum_{u\in out(v)}g_{vu}\leq C_{v},\forall v\in V,\label{eq:MPC3}\end{eqnarray}
 where $U\left(z\right)$ is a twice-differentiable strictly concave
utility function%
\footnote{It might seem unnecessary to involve a strictly concave utility
function in this formulation. The reason is that we later design a
primal-dual algorithm to solve the problem, and using a strictly concave
utility function can avoid its potential instability problem \cite{chen2008ump}.%
}, $\mathbbm{1}_{\{\cdot\}}$ denotes the indicator function. The constraints
in (\ref{eq:MPC1}) describe the flow conservation requirements. The constraints in (\ref{eq:MPC2}) come from the piggybacking
property of information flows. The node upload capacity constraints
are in (\ref{eq:MPC3}). The problem $\mathbf{MP}$ is a convex problem.
All feasible broadcast rates must satisfy the constraints in (\ref{eq:MPC1})-(\ref{eq:MPC3})
and are achievable by using random linear network coding.

\subsection{Algorithm Design via Lagrange Decomposition}

To proceed, we first relax the first set of constraints in \eqref{eq:MPC1}
in problem $\mathbf{MP}$ to obtain a partial Lagrangian as follows:
\begin{align}
  & L(z,\boldsymbol{f,g,\lambda})\nonumber \\
= & U(z)-\sum_{v\in V-\{d\}}\sum_{d\in R}\lambda_{v,d}\left(\sum_{u\in in(v)}f_{uv}^{d}+z\mathbbm{1}_{\left\{ v=s\right\} }-\sum_{u\in out(v)}f_{vu}^{d}\right)\nonumber \\
= & U(z)-\sum_{v\in V}\sum_{d\in R}\lambda_{v,d}\left(\sum_{u\in in(v)}f_{uv}^{d}+z\mathbbm{1}_{\left\{ v=s\right\} }-\sum_{u\in out(v)}f_{vu}^{d}\right),\label{eq:partial.lag}\end{align}
 where $\lambda_{v,d},v\in V-\{d\},d\in R$ are Lagrange multipliers,
$\lambda_{d,d}=0,\forall d\in R$, and $\sum_{u\in in(s)}f_{us}^{d}=0$.

The strong duality holds for problem $\mathbf{MP}$ since the Slater
conditions are satisfied \cite{Boyd04}. Therefore, we can solve problem $\mathbf{MP}$ by finding the saddle points of $L(z,\boldsymbol{f,g,\lambda})$.

Noticing that \begin{align}
\sum_{v\in V}\sum_{d\in R}\lambda_{v,d}\, z\mathbbm{1}_{\left\{ v=s\right\} }=z\sum_{d\in R}\lambda_{s,d}\end{align}
 and \begin{align}
\sum_{v\in V}\sum_{d\in R}\lambda_{v,d}\left(\sum_{u\in in(v)}f_{uv}^{d}-\sum_{u\in out(v)}f_{vu}^{d}\right)=\sum_{d\in R}\sum_{v\in V}\sum_{u\in out(v)}f_{vu}^{d}(\lambda_{u,d}-\lambda_{v,d}),\end{align}
 we can find the saddle points of $L(z,\boldsymbol{f,g,\lambda})$
by solving the following problem successively in $z,\boldsymbol{f},\boldsymbol{g},\boldsymbol{\lambda}$:
\begin{align}
\mathbf{\min_{\boldsymbol{\lambda}\geq0}} & \left(\max_{z\geq0}(U(z)-z\sum_{d\in R}\lambda_{s,d})+\max_{\boldsymbol{f,g}\geq0}\sum_{d\in R}\sum_{v\in V}\sum_{u\in out(v)}f_{vu}^{d}(\lambda_{v,d}-\lambda_{u,d})\right)\label{eq:DP}\\
 & \;\;\;\;\;\;\mbox{s.t.}\;\eqref{eq:MPC2}-\eqref{eq:MPC3}.\nonumber \end{align}

Given $\boldsymbol{\lambda}$ and $z$, we consider the following
scheduling subproblem on $\boldsymbol{f,g}$: \begin{eqnarray}
 & \mathbf{SSP:}\max_{\boldsymbol{f,g}\geq0} & \sum_{d\in R}\sum_{v\in V}\sum_{u\in out(v)}f_{vu}^{d}(\lambda_{v,d}-\lambda_{u,d})\\
 & \mbox{s.t.} & \eqref{eq:MPC2}-\eqref{eq:MPC3}.\nonumber \end{eqnarray}
 The above linear programming problem has a structure that allows
us to solve it distributedly. The first observation is that if
an optimal $\boldsymbol{g}^{*}$ is given, then an optimal $\boldsymbol{f}^{*}$
can be obtained as follows: $\forall u,v\in V,d\in R$, \begin{equation}
\left(f_{vu}^{d}\right)^{*}=\begin{cases}
0, & \mbox{if }\lambda_{v,d}-\lambda_{u,d}\leq0,\\
g_{vu}^{*}, & \mbox{otherwise.}\end{cases}\end{equation}
 As such, it is sufficient to study the following problem in $\boldsymbol{g}$:
\begin{eqnarray}
 & \mathbf{\max_{\boldsymbol{g\ge0}}} & \sum_{v\in V}\sum_{u\in out(v)}g_{vu}\, w_{vu}\\
 & \mbox{s.t.} & \sum_{u\in out(v)}g_{vu}\leq C_{v},\forall v\in V,\nonumber \end{eqnarray}
 where \begin{equation}
w_{vu}\triangleq\sum_{d\in R}[\lambda_{v,d}-\lambda_{u,d}]^{+},\;\forall(u,v)\in E_{f}.\label{eq:back-pressure}\end{equation}
 denotes the aggregate back-pressure between two neighboring nodes
$u$ and $v$, and $[\cdot]^{+}\triangleq\max(\cdot,0)$.

For any $v\in V$, let
\begin{equation}\label{eq:opt.neighbor}
u^{*}(v)\triangleq\arg\max_{u\in out(v)}w_{vu}
\end{equation}
be one of its neighbors with the maximum back-pressure (breaking ties
arbitrarily). Then one optimal solution for problem $\mathbf{SSP}$
is as follows:\begin{equation}
\left(g_{vu}^{d}\right)^{*}=\begin{cases}
C_{v}, & \mbox{if }u=u^{*}(v),\\
0, & \mbox{otherwise,}\end{cases}
\label{eq:g_vu}\end{equation}
 and \begin{equation}
\left(f_{vu}^{d}\right)^{*}=\begin{cases}
0, & \mbox{if }\lambda_{v,d}-\lambda_{u,d}\leq0,\\
g_{vu}^{*}, & \mbox{otherwise.}\end{cases}
\label{eq:f_vu}
\end{equation}
 Given $\boldsymbol{f}^{*}$ and $\boldsymbol{g}^{*}$, primal-dual
algorithms can be designed to adapt $z$ and $\boldsymbol{\lambda}$
to pursue the desired optimal solution.

We summarize the above analysis into a distributed algorithm including
the following components:

\textbf{Primal-dual Rate Control: }we pursue the saddle point in $z$
and $\boldsymbol{\lambda}$ simultaneously as follows: \begin{equation}
\begin{cases}
\dot{z}=\alpha[U^{'}(z)-\sum_{d\in R}\lambda_{s,d}]_{z}^{+},\\
\dot{\lambda}_{v,d}=k_{v,d}\left[\sum_{u\in in(v)}\left(f_{uv}^{d}\right)^{*}+z\mathbbm{1}_{v=s}\right.\\
\;\;\;\;\;\;\;\;\;\;\;\;\;\;\;\;\;\;\;\;\left.-\sum_{u\in out(v)}\left(f_{vu}^{d}\right)_{\lambda_{v,d}}^{*}\right]_{\lambda_{v,d}}^{+}, & \forall v\in V-\{d\},d\in R,\\
\dot{\lambda}_{d,d}=\lambda_{d,d}=0,\;\;\;\forall d\in R,\end{cases}\label{eq:PD}\end{equation}
 where $\alpha$ and $k_{v,d}$ are positive step sizes, and the function
\[
[b]_{a}^{+}=\begin{cases}
\max(0,b), & a\le0,\\
b, & a>0.\end{cases}\]

\textbf{Neighbor Scheduling, Content Scheduling, and Network Coding:}
Every node $v\in V$ maintains a queue storing packets that are intended
for $d$. Whenever a transmission opportunity arises, node $v$ chooses
one neighbor $u^{*}(v)$ with the maximum back-pressure according to \eqref{eq:opt.neighbor}.

If $w_{vu^{*}(v)}>0$, node $v$ sends packets to $u^{*}(v)$ at rate
$C_{v}$. Every output packet is constructed as follows. Node $v$
chooses one packet from the head of each queue of $d$ if $\lambda_{v,d}-\lambda_{u^{*}(v),d}>0$,
and output one random linear combination of these heard-of-queue packets.
If otherwise $w_{vu^{*}(v)}\leq0$ or there is no head-of-line packets
to code, node $v$ does nothing.

We have the following observations.
\begin{itemize}
\item The Lagrangian variable $\lambda_{v,d}$ is proportional to the length
of queue storing packets that are intended for receiver $d$. The
back-pressure $w_{vu}$ measures the aggregate difference in the queues
of all $d\in R$ between $v$ and $u$. The larger the back-pressure is, the
more desperate node $u$ wants to receive data from $v$.
\item Our algorithm can be implemented in a distributed manner. It only
requires nodes to exchange information with its one-hop neighbors,
and thus is robust to peer churn and system dynamics. When a new peer arrives,
it connects to a set of neighbors, assigned by the streaming server or trackers. Then the peer starts exchanging streaming data with them following the strategy defined by our algorithm. When a peer leaves, its neighbors are informed and then close the connections. For the network coding operation, theoretically we need to adjust the size of field where the coding coefficients are chosen to make sure of the decoding probability when the number of nodes changes \cite{all:article:HMKKESL04}, \cite{all:RandomNC:SET03}. While \cite{chou2003practical} and \cite{wang2007r} show that in practice the finite field $\mathbb{F}_{2^8}$ or $\mathbb{F}_{2^{16}}$ is enough to have a sufficiently high decoding probability. Therefore, only local
configuration changes corresponding to dynamics, which is easy to implement
compared to centralized algorithms where typically global information is needed for whole configuration change (e.g., spanning trees reconstruction in spanning tree based solutions).
\item Although our algorithm is designed for P2P broadcast scenarios, it
also works for P2P multicast scenarios where helper nodes exist. The
helper nodes simply also perform the operations described in
\eqref{eq:g_vu}-\eqref{eq:PD}. Our algorithm can be considered as
the extension of the algorithm in~\cite{ho2009dynamic} from
link-capacity-limited underlay networks to node-capacity-limited
overlay networks.
\end{itemize}

The following theorem characterize the convergence of the proposed algorithm.
\begin{theorem}\label{thm:converge}
The algorithm in \eqref{eq:g_vu}-\eqref{eq:PD} converges to the optimal solution of problem $\mathbf{MP}$
globally asymptotically in time.
\end{theorem}
The proof utilizes standard Lyapunov arguments and a Lyapunov function for primal-dual algorithm, similar to those used
in~\cite{chen2008ump}, \cite{all:mathematics.cong.ctrl}. The proof is relegated to Appendix \ref{sec:proof_converge}.

\textbf{Remark:} We derive our algorithm and prove its convergence based on a fluid model formulation.
It is also possible to obtain a similar back-pressure based distributed algorithm with
packet-level dynamics taken into account and prove its stability,  following
a set of Lyapunov drift arguments elaborated in \cite{georgiadis2006resource}.

%

\section{The Proposed Distributed Topology Hopping Algorithm}

\label{sec:maf}

We recently proposed in \cite{MA:CLSC10} to use Markov chain as a principled approach in designing distributed algorithms for solving combinatorial network problems approximately. In particular, we show one can design distributed algorithms for a combinatorial network optimization problem in the following way. First, construct a special class of Markov chains with problem-specific steady-state distribution. Second, search for a Markov chain in this class that allows distributed implementation. If such Markov chain can be found, which is usually challenging and problem-specific, the distributed implementation directly yields a distributed algorithm for the problem.

In this paper, we follow the framework from \cite{MA:CLSC10} and design a distributed topology hopping algorithm for our problem \eqref{MWC}. There are two steps in designing
our algorithm under the Markov approximation framework \cite{MA:CLSC10}:
log-sum-exp approximation and constructing problem-specific Markov chains that allows distributed implementation.

%
{}

\subsection{Log-Sum-Exp Approximation}

First, the maximum broadcast rate can be approximated by a log-sum-exp
function as follows: \begin{align}
\max_{f\in\mathcal{F}}\; x_{f}\,\approx\,\frac{1}{\beta}\log\left[\sum_{f\in\mathcal{F}}\exp\left(\beta x_{f}\right)\right],\label{appro1}\end{align}
 where $\beta$ is a positive constant. Let $|\mathcal{F}|$ denote
the size of the set $\mathcal{F}$, then the approximation accuracy
is known as follows \cite{MA:CLSC10}: \begin{align}
0\le\frac{1}{\beta}\log\left[\sum_{f\in\mathcal{F}}\exp\left(\beta x_{f}\right)\right]-\max_{f\in\mathcal{F}}\; x_{f}\le\frac{1}{\beta}\log|\mathcal{F}|.\label{aa1}\end{align}

As $\beta$ approaches infinity, the approximation gap approaches
zero. As discussed in \cite{MA:CLSC10}, however, usually $\beta$ should not
take too large values as there are practical constraints or convergence
rate concerns in the algorithm design afterwards.

To better understand the log-sum-exp approximation, we associate with each configuration $f\in\mathcal{F}$
a probability $p_{f}$. Consider the following problem
\begin{eqnarray}
\mathbf{MRC-EQ:}\max_{\boldsymbol{p\geq0}} &  & \sum_{f\in\mathcal{F}}p_{f}x_{f}\\
\mbox{s.t.} &  & \sum_{f\in\mathcal{F}}p_{f}=1.\end{eqnarray}
Its optimal value is $\max_{f\in \mathcal{F}} x_f$ and is obtained by setting the probability corresponding to one of the best configurations
to be one and the rest probabilities to be zero. Hence, problem $\mathbf{MRC-EQ}$
is equivalent to the original problem $\mathbf{MRC}$.

On the other hand, according to \cite{MA:CLSC10} we have the following
observations.
\begin{theorem}[cf. \cite{MA:CLSC10}] The optimal value of the following
optimization problem \begin{align}
\mathbf{MRC-\beta:} & \max_{\boldsymbol{p}\geq0}\sum_{f\in\mathcal{F}}p_{f}x_{f}-\frac{1}{\beta}\sum_{f\in\mathcal{F}}p_{f}\log p_{f}\;\;\;\;\;\;\label{conju}\\
 & \mbox{s.t.}\;\;\sum_{f\in\mathcal{F}}p_{f}=1\label{conju_cs}\end{align}
is given by $\frac{1}{\beta}\log\left[\sum_{f\in\mathcal{F}}\exp\left(\beta x_{f}\right)\right]$.
The optimal solution of problem $\mathbf{MRC-\beta}$ is given by
\begin{align}
p_{f}^{*}(\boldsymbol{x})=\frac{\exp\left(\beta x_{f}\right)}{\sum\limits _{f'\in\mathcal{F}}\exp\left(\beta x_{f'}\right)},\;\forall f\in\mathcal{F}.\label{dist_no_error}\end{align}
 \end{theorem}

As such, by the log-sum-exp approximation in (\ref{appro1}),
we obtain an approximate version of the maximum broadcast rate problem
$\mathbf{MRC}$, off by an \emph{entropy }term $-\frac{1}{\beta}\sum_{f\in\mathcal{F}}p_{f}\log p_{f}$.
If we can time-share among different configurations according
to the optimal solution $p_{f}^{*}(\boldsymbol{x})$ in \eqref{dist_no_error},
then we can solve the problem $\mathbf{MRC}$ approximately and obtain
a close-to-optimal broadcast rate.

\subsection{Markov Chain Guided Algorithm Design}

We design a Markov chain with a state space being the set of all feasible peering configurations $\mathcal{F}$ and has a stationary distribution as $p_{f}^{*}(\boldsymbol{x})$ in \eqref{dist_no_error}. We implement the Markov chain to guide the system to optimize the configuration. As the system hops among configurations, the Markov chain converges and the configurations are time-shared according to the desired distribution $p_{f}^{*}(\boldsymbol{x})$.

The key lies in designing such Markov chain that allows distributed
implementation. Since $p_{f}^{*}(\boldsymbol{x})$ in \eqref{dist_no_error}
is product-form, it suffices to focus on designing time-reversible Markov chains
\cite{MA:CLSC10}.

Let $f,f'\in\mathcal{F}$ be two states of Markov chain, and denote
$q_{f,f^{'}}$ as the transition rate from state $f$
to $f^{'}$. We have two degrees of freedom in designing a time-reversible
Markov chain:
\begin{itemize}
\item \textbf{The state space structure}: we can add or
cut direct transitions between any two states, given that the state
space remains connected and any two states are reachable from each
other.
\item \textbf{The transition rates}: we can explore various options
in designing $q_{f,f^{'}}$, given that the detailed balance
equation is satisfied, i.e., \begin{equation}
p_{f}^{*}(\boldsymbol{x})q_{f,f^{'}}=p_{f}^{*}(\boldsymbol{x})q_{f^{'},f},\;\forall f,f'\in\mathcal{F}.\label{eq:detailed.balance.eq}\end{equation}
Satisfying the above equations guarantees the designed Markov chain has the desired stationary distribution
as in \eqref{dist_no_error}.
\end{itemize}

Recall that for a node $v\in V$, the set of its neighbors under configuration
$f$ is denoted by $N_{v,f}$. We call node in $N_{v,f}$ $v$'s in-use neighbor and node in $N_{v}\backslash N_{v,f}$ $v$'s not-in-use neighbor. For the ease of explanation,
we further define $\mathcal{N}_{f}$ as the set of all the node-pairs under $f$, i.e., $\mathcal{N}_{f}=\{\{v,u\}, \forall v\in V,u\in N_{v,f}\}$. Note we do not differentiate node pairs $\{u, v\}$ and $\{v, u\}$.
As an example, for the peering configuration $f$ shown in Fig.~\ref{fig:pc}(b), $\mathcal{N}_{f}$ is given by $\{\{s,1\},\{s,2\},\{s,4\},\{1,2\},\{1,4\},\{2,3\}, \{3,4\}\}$.

In our Markov chain design, we first specify its state space structure
as follows: we set the transition rate $q_{f,f'}$ to be zero, unless $f$ and $f'$ satisfy that
$|\mathcal{N}_{f}\backslash\mathcal{N}_{f'}|=1$ or $|\mathcal{N}_{f'}\backslash\mathcal{N}_{f}|=1$.
In other words, we only allow direct transitions between two configurations
if such transitions correspond to a single node adding a new node in its in-use
neighbor set or removing one in-use neighbor from its in-use neighbor set.

Second, given the state space structure of Markov chain, we design
the transition rates to favor distributed implementation
while satisfying the detailed balance equation in (\ref{eq:detailed.balance.eq}).

One possible option is to set $q_{f,f^{'}}$ to be $\exp^{-1}(\beta x_{f})$.
One way to implement this option is for every node to generate a timer according to its
\emph{measured} receiving rate and counts down accordingly. When the
timer expires, the dedicated node performs the neighbor swapping and resets its timer. As
simple as the implementation may sound, this option is expensive to implement. Once
the peering configuration changes, the system needs to notify all
the nodes to measure the new receiving rate and reset their timers
accordingly. It is not clear how to implement such system-wide notification
in a low-overhead manner.

In this paper, we design $q_{f,f^{'}}$ and
$q_{f',f}$ as follows:
\begin{equation}
q_{f,f^{'}}=\frac{1}{\exp(\tau)}\frac{\exp(\beta x_{f^{'}})}{\exp(\beta x_{f})+\exp(\beta x_{f^{'}})}\label{eq:transition rate}
\end{equation}
and
\begin{equation}
q_{f',f}=\frac{1}{\exp(\tau)}\frac{\exp(\beta x_{f})}{\exp(\beta x_{f})+\exp(\beta x_{f^{'}})},
\end{equation}
where $\tau$ is a constant. It is straightforward to verify that
detailed balance equation is satisfied. As will be clear in the next subsection, our choices of transition rates
do not require coordination or notification among peers in its implementation.

\subsection{Distributed Implementation\label{sub:Distributed-Implementation}}

One distributed implementation of our designed Markov chain is briefly
described as follows.
\begin{itemize}
\item \textbf{Initialization:} Each peer $v\in V$ randomly selects
neighbors from its neighbor list $N_{v}$ under the node degree bound and builds connections with these selected neighbors.
\item \textbf{Step 1:} Let $f$ denote the current configuration. Each node
$v\in V$ generates an exponentially distributed random number independently
with mean $\frac{2\exp(\tau)}{|N_{v}|}$,
and counts down according to this number.
\item \textbf{Step 2:} When the count-down expires, node $v$ measures its
current receiving rate as an estimate of the broadcast rate $x_{f}$.
Then with probability $\frac{|N_{v,f}|}{|N_{v}|}$ node $v$ goes to the \textbf{Step 2a};
with probability $\frac{|N_{v}|-|N_{v,f}|}{|N_{v}|}$, node $v$ goes to the \textbf{Step 2b};
    \begin{itemize}
        \item \textbf{Step 2a:} Node $v$ randomly selects one in-use neighbor
        in $N_{v,f}$ and removes it from $N_{v,f}$. Under the new peering
        configuration $f^{'}$, node $v$ measures its receiving rate as an estimate of
        $x_{f^{'}}$. With the estimates of $x_{f}$ and $x_{f^{'}}$, peer $v$ stays in the
        new configuration $f^{'}$ with probability $\frac{\exp(\beta x_{f^{'}})}{\exp(\beta x_{f})+\exp(\beta x_{f^{'}})}$, and switches back to $f$ with probability $1-\frac{\exp(\beta x_{f^{'}})}{\exp(\beta x_{f})+\exp(\beta x_{f^{'}})}$. Node $v$ then repeats \textbf{Step 1}.
        \item \textbf{Step 2b:} Node $v$ randomly selects one not-in-use neighbor
        in $N_{v}\backslash N_{v,f}$. If the node degree of the selected not-in-use node is equal to
        the bound or $v$'s node degree is equal to the bound, node $v$ jumps back to
        \textbf{Step 1} immediately. Otherwise,
        node $v$ adds this selected node into $N_{v,f}$. Under the new peering
        configuration $f^{'}$, node $v$ measures its receiving rate as an estimate of
        $x_{f^{'}}$. With the estimates of $x_{f}$ and $x_{f^{'}}$, peer $v$ stays in the
        new configuration $f^{'}$ with probability $\frac{\exp(\beta x_{f^{'}})}{\exp(\beta x_{f})+\exp(\beta x_{f^{'}})}$, and switches back to $f$ with probability $1-\frac{\exp(\beta x_{f^{'}})}{\exp(\beta x_{f})+\exp(\beta x_{f^{'}})}$. Node $v$ then repeats \textbf{Step 1}.
    \end{itemize}
\end{itemize}

It is straightforward to summarize the above implementation into a distributed algorithm that runs on individual nodes and utilizes only the measurement from their one-hop neighbors. The correctness of the implementation is shown as follows:
\begin{proposition}
\label{implement_mc} The implementation in fact realizes a time-reversible
Markov chain with stationary distribution in \eqref{dist_no_error}.
\end{proposition}
The proof is relegated to Appendix \ref{sec:proof_mc}.

\textbf{Remarks:} a) In \textbf{Step 1}, the generation of count-down
timers does not depend on the receiving rate, thus the system does
not need to notify the nodes about changes of peering configurations.
b) With the above implementation, the system hops towards configurations
with better broadcast rate probabilistically. For example, if $x_{f'}>x_{f}$,
then the system will be more likely to stay in configuration $f'$
than in $f$, and vice versa. c) With large values of $\beta$, the system
hops towards better configurations more greedily. However,
this may as well lead to the system getting trapped in locally optimal configurations.
Hence there is a trade-off to consider when setting the value of $\beta$.
Moreover, the value of $\beta$ also affects the convergence rate
of the time-reversible Markov chain to the desired stationary distribution.
It is worth future investigation to further understand the impact
of $\beta$. d) In the presence of peer dynamics, our algorithm incurs only simple actions based on local information. When a new peer arrives, a neighbor set and a neighbor list are assigned
to it. The peer builds connections with the nodes in the neighbor set. Then the peer starts counting down as \textbf{Step 1} and follows the strategy of our algorithm. When a peer leaves, we just eliminate it from the neighbor list of its previous neighbors and end up connections.

\section{Convergence Properties of Overall Solution}

\label{sec:bp+ma}

\begin{algorithm}[hbt!]
\caption{Broadcasting Algorithm}
\label{alg:bp}
\begin{algorithmic}[1]
\STATE The following procedure runs on each individual node independently.
\STATE For the source $s$ and each time slot,
\STATE $\;\;\;$$x\leftarrow \left[x+ \alpha(U^{'}(x)-\sum_{d\in R}\lambda_{s,d})\right]^{+}$
\STATE For each node $v\in V$ and each time slot,
\STATE $w^{*}\leftarrow 0$
\FOR {$u\in out(v)$}
   \FOR {for $d\in R$}
   \STATE {$w_{vu}\leftarrow w_{vu}+\mbox{max}(\lambda_{v,d}-\lambda_{u,d}, 0)$}
   \ENDFOR
   \IF {$w_{vu} > w^{*}$}
   \STATE {$w^{*}\leftarrow w_{vu}$}
   \STATE {$u^{*}\leftarrow u$}
   \ENDIF
\ENDFOR
\IF {$w_{vu^{*}} > 0$}
   \FOR {$d\in R$}
      \IF {$\lambda_{v,d}-\lambda_{u^{*},d} > 0$}
      \STATE {$f_{vu^{*}}^{d}\leftarrow C_v$}
      \ENDIF
   \ENDFOR
\ENDIF
\FOR {$d\in R$}
\STATE {$\lambda_{v,d}\leftarrow \left[\lambda_{vd}+k_{v,d}(\sum_{u\in in(v)}f_{uv}^{d}-\sum_{u\in out(v)}f_{vu}^{d})\right]^{+}$}
\ENDFOR
\end{algorithmic}
\end{algorithm}

\begin{algorithm}[hbt!]
\caption{Topology Hopping Algorithm}
\label{alg:ma}
\begin{algorithmic}[1]
\STATE The following procedure runs on each individual node independently. We focus on a particular node $v\in V$.
\STATE \textbf{procedure} Initialization
\begin{itemize}
\item Initialize $N_v$, $B_v$; randomly connects to peers from $N_v$ under the degree bound.
\item Generate a timer that follows exponential distribution with mean equal to $2\exp(\tau)/(|N_{v}|)$ and begin counting down.
\end{itemize}
\STATE \textbf{end procedure}
\STATE {$\,$}
\STATE {When the timer expires, invoke the procedure Transition.}
\STATE \textbf{procedure} Transition
\STATE $\quad$With probability $\frac{|N_{v,f}|}{|N_{v}|}$,\\
\STATE $\quad$$\quad$$N_o\leftarrow N_{v,f}$;\\
\STATE $\quad$$\quad$randomly remove one in-use neighbor from $N_{v,f}$;\\
\STATE $\quad$$\quad$$x_{f^{'}}\leftarrow\sum_{u\in in(v)}f_{uv}^{v}$;\\
\STATE $\quad$$\quad$$N_{v,f}$ $\leftarrow$ $N_o$ with probability\\
       $\quad$$\quad$$1-{\exp(\beta x_{f^{'}})}/{\left(\exp(\beta x_f)+\exp(\beta x_{f^{'}})\right)}$;\\
\STATE $\quad$$\quad$refresh the timer and begin counting down;\\
\STATE $\quad$With probability $1-\frac{|N_{v,f}|}{|N_{v}|}$,\\
\STATE $\quad$$\quad$$N_o\leftarrow N_{v,f}$;\\
\STATE $\quad$$\quad$randomly add one not-in-use neighbor $v^{'}$ in\\
       $\quad$$\quad$$N_{v}\backslash N_{v,f}$ to $N_{v,f}$;\\
\STATE $\quad$$\quad$\textbf{if} $|N_{v,f}|=B_{v}$ or $|N_{v^{'},f}|=B_{v^{'}}$\\
\STATE $\quad$$\quad$$\quad$refresh the timer and begin counting down;\\
\STATE $\quad$$\quad$\textbf{end if}\\
\STATE $\quad$$\quad$$x_{f^{'}}\leftarrow\sum_{u\in in(v)}f_{uv}^{v}$;\\
\STATE $\quad$$\quad$$N_{v,f}$ $\leftarrow$ $N_o$ with probability\\
       $\quad$$\quad$$1-{\exp(\beta x_{f^{'}})}/{\left(\exp(\beta x_f)+\exp(\beta x_{f^{'}})\right)}$;\\
\STATE $\quad$$\quad$refresh the timer and begin counting down;\\
\STATE \textbf{end procedure}
\end{algorithmic}
\end{algorithm}

We have designed the distributed broadcasting algorithm in Section
\ref{sec:streaming_rate} and the Markov chain guided topology hopping
algorithm in Section \ref{sec:maf}. The pseudocodes of each algorithm
are shown in Algorithm~\ref{alg:bp} and Algorithm~\ref{alg:ma} respectively.
Both algorithms are simple to implement, run on each individual node,
and only require nodes to exchange information with their neighbors.

If the broadcasting algorithm converges instantaneously, i.e., time-scale
separation assumption holds, then we can obtain the accurate value
of $x_{f}$ for any configuration $f\in\mathcal{F}$. Transiting based
on the accurate $x_{f}$, the designed Markov chain will converges
to the desired stationary distribution in \eqref{dist_no_error}.
Hence by operating these two algorithms in tandem, we obtain a close-to-optimal
broadcast rate under arbitrary node degree bounds, and over arbitrary
overlay graph. The optimality gap is characterized in (\ref{aa1}).

In practice, however, it is possible to obtain only an inaccurate
measurement or estimate of $x_{f}$. These inaccuracies root in two
sources. One is the noisy measurements of the maximum broadcast rates
given the configuration. The other is the fast state transition of
Markov chain, i.e., the Markov chain transits before the underlying
broadcasting algorithm converges and thus it transits based on inaccurate
observations of the broadcast rates.

Consequently, the topology hopping Markov chain may \emph{not} converge
to the desired stationary distribution $p_{f}^{*}(\boldsymbol{x})$.
This observation motivates our following study on the convergence
of Markov chain in the presence of inaccurate transition rates.

For each configuration $f\in\mathcal{F}$ with broadcast rate $x_{f}$,
we assume its corresponding inaccurate observed rate belongs to the
bounded region $\left[-\Delta_{f},\Delta_{f}\right]$. $\Delta_{f}$
is the inaccuracy bound and can be different for different $f$.

For easy explanation of our approach, we further assume the observed
broadcast rate for configuration $f$ only takes one of the following
$2n_{f}+1$ discrete values: \[
\left[x_{f}-\Delta_{f},\ldots,x_{f}-\frac{1}{n_{f}}\Delta_{f},x_{f},x_{f}+\frac{1}{n_{f}}\Delta_{f},\ldots,x_{f}+\Delta_{f}\right],\]
 where $n_{f}$ is a positive constant. Further, with probability
$\eta_{j,f}$, the observed broadcast rate takes value
$x_{f}+\frac{j}{n_{f}}\Delta_{f}$, $\forall
j\in\{-n_{f},\ldots,n_{f}\}$ and
$\sum_{j=-n_{f}}^{n_{f}}\eta_{j,f}=1$.

With the inaccurate observed broadcast rates, the topology hopping
behaves as follows. Suppose the current configuration is $f$ and
the observed broadcast rate is $x_{f}+\frac{j}{n_{f}}\Delta_{f}$,
where $j\in\{-n_{f},\ldots,n_{f}\}$. After some count-down process,
the system hops to a new configuration $f'$ and probes its broadcast
rate. In configuration $f'$, the broadcast rate is observed as $x_{f'}+\frac{j'}{n_{f'}}\Delta_{f'},j'\in\{-n_{f'},\ldots,n_{f'}\}$.
The system stays in the new configuration $f'$ with probability \[
\frac{\exp(\beta(x_{f'}+\frac{j'}{n_{f'}}\Delta_{f'}))}{\exp(\beta(x_{f'}+\frac{j'}{n_{f'}}\Delta_{f'}))+\exp(\beta(x_{f}+\frac{j}{n_{f}}\Delta_{f}))},\]
and switches back to configuration $f$ with probability \[
1-\frac{\exp(\beta(x_{f'}+\frac{j'}{n_{f'}}\Delta_{f'}))}{\exp(\beta(x_{f'}+\frac{j'}{n_{f'}}\Delta_{f'}))+\exp(\beta(x_{f}+\frac{j}{n_{f}}\Delta_{f}))}.\]
By arguments similar to the proof of Proposition~$1$, the transition
rate from configuration $f$ with broadcast rate $x_{f}+\frac{j}{n_{f}}\Delta_{f}$
to configuration $f'$ with broadcast rate $x_{f'}+\frac{j'}{n_{f'}}\Delta_{f'}$
is given by \begin{align}
\frac{\eta_{j',f'}}{\exp(\tau)}\cdot\frac{\exp(\beta(x_{f'}+\frac{j'}{n_{f'}}\Delta_{f'}))}{\exp(\beta(x_{f'}+\frac{j'}{n_{f'}}\Delta_{f'}))+\exp(\beta(x_{f}+\frac{j}{n_{f}}\Delta_{f}))}.\end{align}

\begin{figure}
\centering \includegraphics[scale=0.4]{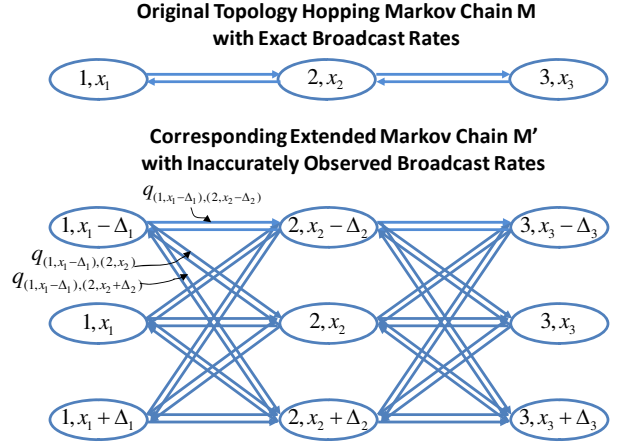} \centering
\caption{An example of the original three-state topology hopping Markov chain and the extended Markov chain. M is the original topology
hopping Markov chain with accurate broadcast rates.
M' is the corresponding extended Markov chain with inaccurate broadcast
rate observations. For each configuration $f\in\{1,2,3\}$,
the observed broadcast rate takes values $x_{f}-\Delta_{f}$, $x_{f}$,
$x_{f}+\Delta_{f}$ with probability $\eta_{-1,f},\eta_{0,f}$ and $\eta_{1,f}$
respectively. The transition rates are assigned according to \eqref{tran_rate_mc1} and \eqref{tran_rate_mc2}.
}

\label{mc_inexact}
\end{figure}

We construct a Markov chain to capture and study the above topology
hopping behavior. In this Markov chain, a state is associated with
a configuration and an observed broadcast rate. Given any configuration
$f\in\mathcal{F}$ and its corresponding $x_{f}$, there are
$2n_{f}+1$ states in the extended Markov chain: $\left(f,x_{f}+\frac{j}{n_{f}}\Delta_{f}\right),j\in\{-n_{f},\ldots,n_{f}\}$.
Further, Given direct transitions between configuration $f$ and $f'$
in the original topology hopping Markov chain, there are direct transitions
between states $(f,x_{f}+\frac{j}{n_{f}}\Delta_{f})$ and $(f',x_{f'}+\frac{j'}{n_{f'}}\Delta_{f'})$
($\forall j\in\{-n_{f},\ldots,n_{f}\},j'\in\{-n_{f'},\ldots,n_{f'}\}$)
in the corresponding new Markov chain. The corresponding transition
rates are shown as follows: \begin{align}
 & q_{(f,x_{f}+\frac{j}{n_{f}}\Delta_{f}),(f',x_{f'}+\frac{j'}{n_{f'}}\Delta_{f'})}\nonumber \\
= & \frac{\eta_{j',f'}}{\exp(\tau)}\cdot\frac{\exp(\beta(x_{f'}+\frac{j'}{n_{f'}}\Delta_{f'}))}{\exp(\beta(x_{f'}+\frac{j'}{n_{f'}}\Delta_{f'}))+\exp(\beta(x_{f}+\frac{j}{n_{f}}\Delta_{f}))}\label{tran_rate_mc1}\end{align}
 and \begin{align}
 & q_{(f',x_{f'}+\frac{j'}{n_{f'}}\Delta_{f'}),(f,x_{f}+\frac{j}{n_{f}}\Delta_{f})}\nonumber \\
= &
\frac{\eta_{j,f}}{\exp(\tau)}\cdot\frac{\exp(\beta(x_{f}+\frac{j}{n_{f}}\Delta_{f}))}{\exp(\beta(x_{f}+\frac{j}{n_{f}}\Delta_{f}))+\exp(\beta(x_{f'}+\frac{j'}{n_{f'}}\Delta_{f'}))}\label{tran_rate_mc2},\end{align}
where $\sum_{j=-n_{f}}^{n_{f}}\eta_{j,f}=1$ and
$\sum_{j'=-n_{f'}}^{n_{f'}}\eta_{j',f'}=1$. This new Markov chain
can be thought as an extended version of the original topology
hopping Markov chain. As an example, an extended Markov chain is
shown and explained in Fig. \ref{mc_inexact}.

The extended Markov chain has a unique stationary distribution since
it is irreducible and only has a finite number of states. We can study
the impact of inaccurate broadcast rates by comparing the stationary
configuration distribution of the new Markov chain and that
of the original topology hopping Markov chain.

We denote the stationary distribution of the \emph{states} in the
new Markov chain by \begin{align}
\boldsymbol{\tilde{p}}\triangleq[\tilde{p}_{f,x_{f}+\frac{j}{n_{f}}\Delta_{f}},j\in\{-n_{f},\ldots,n_{f}\},f\in\mathcal{F}].\end{align}
We also denote $\boldsymbol{\bar{p}}:[\bar{p}_{f}(\boldsymbol{x}),f\in\mathcal{F}]$
as the stationary distribution of the \emph{configurations} in the
extended Markov chain. Given a configuration $f\in\mathcal{F}$, there
are $2n_{f}+1$ states associated with $f$ in the extended Markov
chain. We have \begin{align}
\bar{p}_{f}(\boldsymbol{x})=\sum_{j\in\{-n_{f},\ldots,n_{f}\}}\tilde{p}_{f,x_{f}+\frac{j}{n_{f}}\Delta_{f}},\forall f\in\mathcal{F}.\end{align}

Recall that the stationary distribution of the configurations for
the original topology hopping Markov chain is $\boldsymbol{p^{*}}:[p_{f}^{*}(\boldsymbol{x}),f\in\mathcal{F}]$.
We use the total variance distance \cite{Diaconis91} to quantify
the difference between $\boldsymbol{p^{*}}$ and $\boldsymbol{\bar{p}}$,
as \begin{align}
d_{TV}(\boldsymbol{p^{*}},\boldsymbol{\bar{p}})\triangleq\frac{1}{2}\sum\limits _{f\in\mathcal{F}}|p_{f}^{*}-\bar{p}_{f}|.\end{align}

We have the following result: \begin{theorem} \label{error_impact}
Let $\Delta_{\max}=\max_{f\in\mathcal{F}}\Delta_{f}$, and $x_{\max}=\max_{f\in\mathcal{F}}x_{f}$.
The $d_{TV}(\boldsymbol{p^{*}},\boldsymbol{\bar{p}})$ are bounded
as follows: \begin{align}
0\le d_{TV}(\boldsymbol{p^{*}},\boldsymbol{\bar{p}})\le1-\exp\left(-2\beta\Delta_{\max}\right).\label{upper_bound}\end{align}
 Further, the optimality gap in broadcast rates $|\boldsymbol{p^{*}}\boldsymbol{x}^T-\boldsymbol{\bar{p}}\boldsymbol{x}^T|$ is bounded as below:
\begin{equation}
|\boldsymbol{p^{*}}\boldsymbol{x}^T-\boldsymbol{\bar{p}}\boldsymbol{x}^T|\leq 2x_{\max}(1-\exp(-2\beta\Delta_{\max})).
\end{equation}
\end{theorem}
The proof is relegated to Appendix\ref{sec:error_impact}.

\textbf{Remarks:} a) The upper bound on
$d_{TV}(\boldsymbol{p^{*}},\boldsymbol{\bar{p}})$ shown in
\eqref{upper_bound} is general, as it is independent of the number
of configurations $|\mathcal{F}|$, the values of $n_{f}$, and the
distributions of inaccurate observed rates
$\eta_{j,f}\;\left(-n_{f}\leq j\leq n_{f},f\in\mathcal{F}\right)$.
b) The upper bound on
$d_{TV}(\boldsymbol{p^{*}},\boldsymbol{\bar{p}})$ shown in
\eqref{upper_bound} decreases exponentially with the worst
inaccuracy bound $\Delta_{\max}$ decreasing. c) It would be
interesting to explore a tighter upper bound on
$d_{TV}(\boldsymbol{p^{*}},\boldsymbol{\bar{p}})$ than the one in
\eqref{upper_bound}.

\section{Performance Evaluation}

\label{sec:simu}

We implement a packet-level simulator to our proposed solutions and use this simulator to evaluate the performance of our solutions.

\subsection{Settings}

In our simulations, time is chopped into slots of equal length, and we adopt three different settings. In Setting I, we set the total number of nodes to be $100$, and assign the node upload capacities randomly according to the distribution
in Table \ref{tab:distrib}, which is obtained from the uplink bandwidth statistics of Internet hosts~\cite{vodprofitable}. We set the source's upload capacity to be $768$ kbps; with this upload capacity, source is not the broadcast bottleneck~\cite{all:Mutualcast:LPZ05,all:P2PStreaming:KLR.07}.

Setting II is the same as Setting I, except we set the total number of nodes to be $10$.

\begin{figure}[hbt!]
\centering
\subfigure{\includegraphics[width=0.35\textwidth]{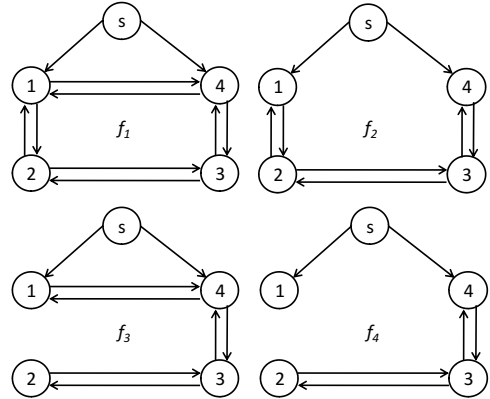}}
\caption{Peering configurations under Setting III. For the ease of illustration, we only allow node $1$ to add or remove neighbors between nodes $2$ and $4$. The rest nodes keep their neighbors fixed.}
\label{fig:4_config}
\end{figure}

In Setting III, there are $4$ different peering configurations as shown in Fig.~\ref{fig:4_config}. Every node has a unit capacity. Under configuration $f_1$, $f_2$ and $f_3$ the maximum broadcast rate is $1$, and under configuration $f_4$ the maximum broadcast rate is $0.5$.

When running our network coding based broadcasting algorithm, we set the updating step size of $z$ and $\boldsymbol{\lambda}$ to be $0.1$ and $0.00005$ respectively. These parameters are empirically chosen to obtain smooth algorithm updating and small errors.

In our simulations, we assign node degree bounds in the following two ways. The first is to set identical bound on each node's node degree. The second is to set degree bound proportional to the node's upload capacity. This is based on the empirical observations that nodes with high upload capacities usually have more system resource (e.g., memory and CPU power) than nodes with low upload capacities. With more system resource, nodes can maintain more concurrent connections, thus have larger node degree bounds. In our second degree bounds assignment, nodes set their node degree bounds proportional to the ratio between their upload capacities and $64$ kbps. In particular, nodes with $64$ kbps have a degree bound of $2$, and nodes with $128$ kbps have a degree bound of $4$, etc.

We carry out two sets of simulations. First, we evaluate the performance of our distributed broadcasting algorithm under Setting I and II. Second, we evaluate the overall performance when we combine the topology hopping algorithm and the broadcasting algorithm under Setting I and III. In these two sets of simulations, we also compare the performance under the two degree bounds assignments explained in the previous paragraph.

\begin{table}
\caption{Peer upload capacity distribution}

\label{tab:distrib}

\centering{}\begin{tabular}{|c|c|c|c|c|c|}
\hline
Upload Capacity (kbps)  & 64  & 128  & 256  & 384  & 768 \tabularnewline
\hline
Fraction (\%)  & 2.8  & 14.3  & 4.3  & 23.3  & 55.3 \tabularnewline
\hline
\end{tabular}
\end{table}

\begin{figure*}[hbt!]
\centering \mbox{ \subfigure[]{\includegraphics[width=0.25\textwidth]{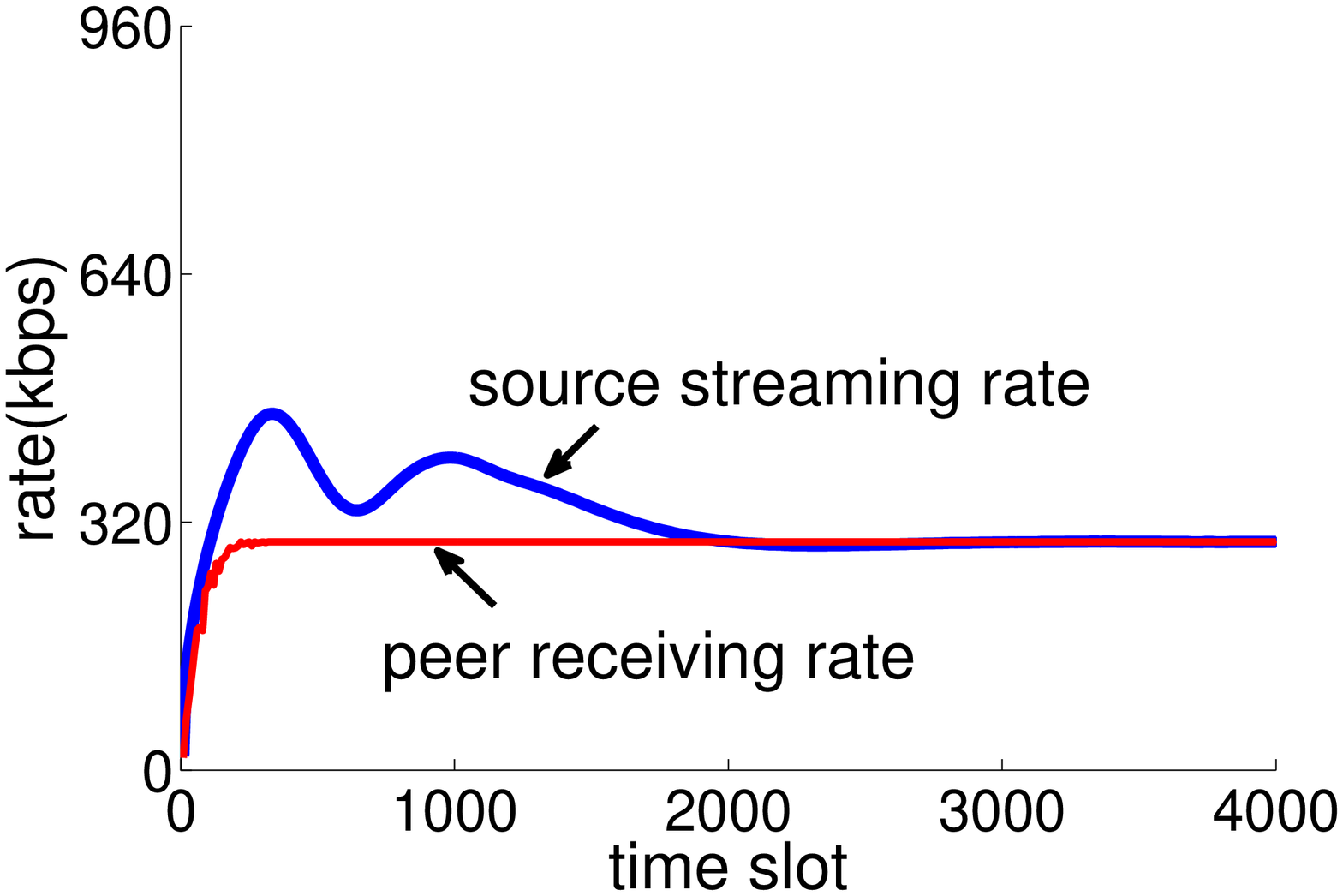}\label{fig:bp.10.bound3}}
\subfigure[]{\includegraphics[width=0.25\textwidth]{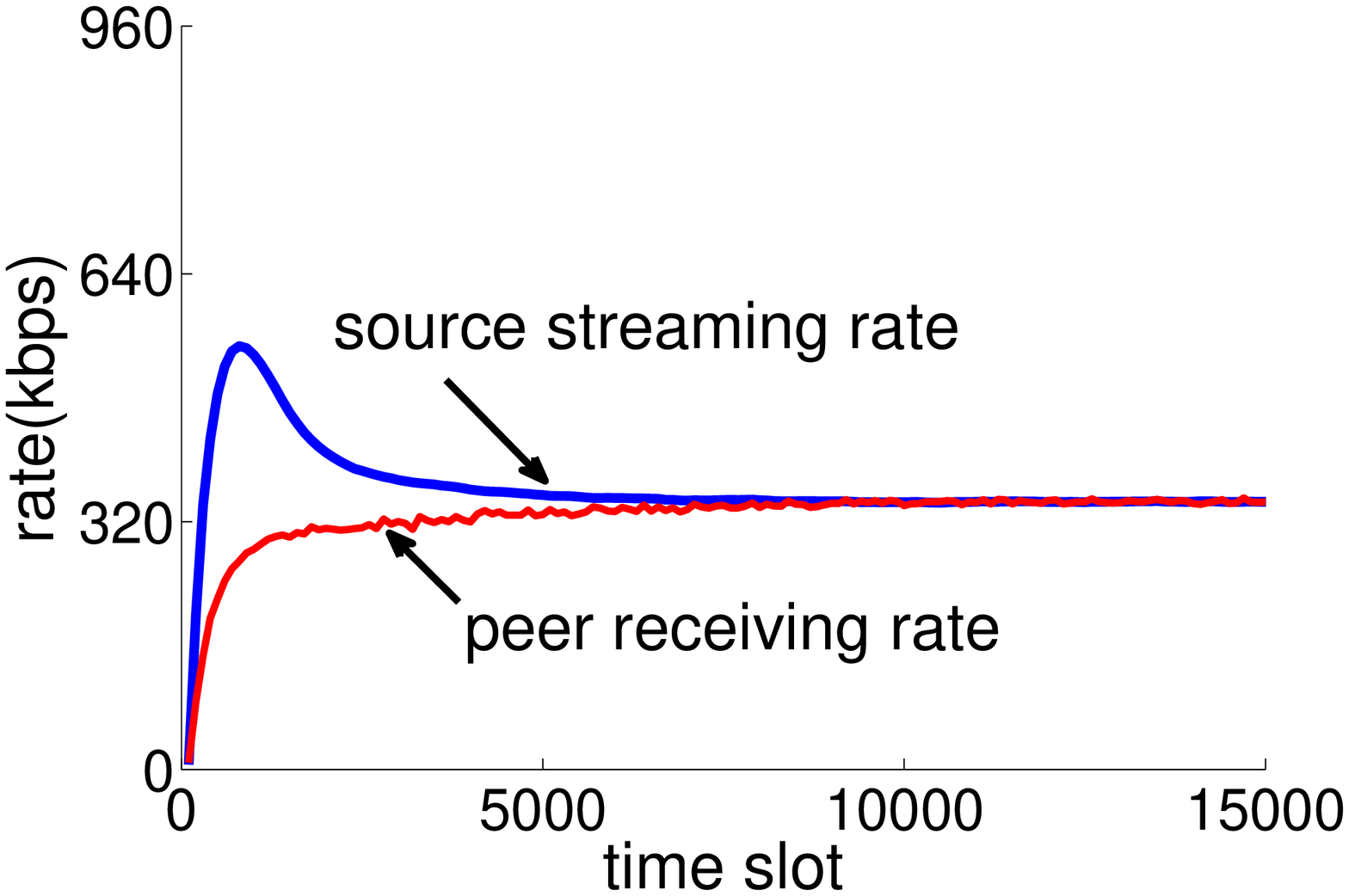}\label{fig:bp.bound3}}
\subfigure[]{\includegraphics[width=0.25\textwidth]{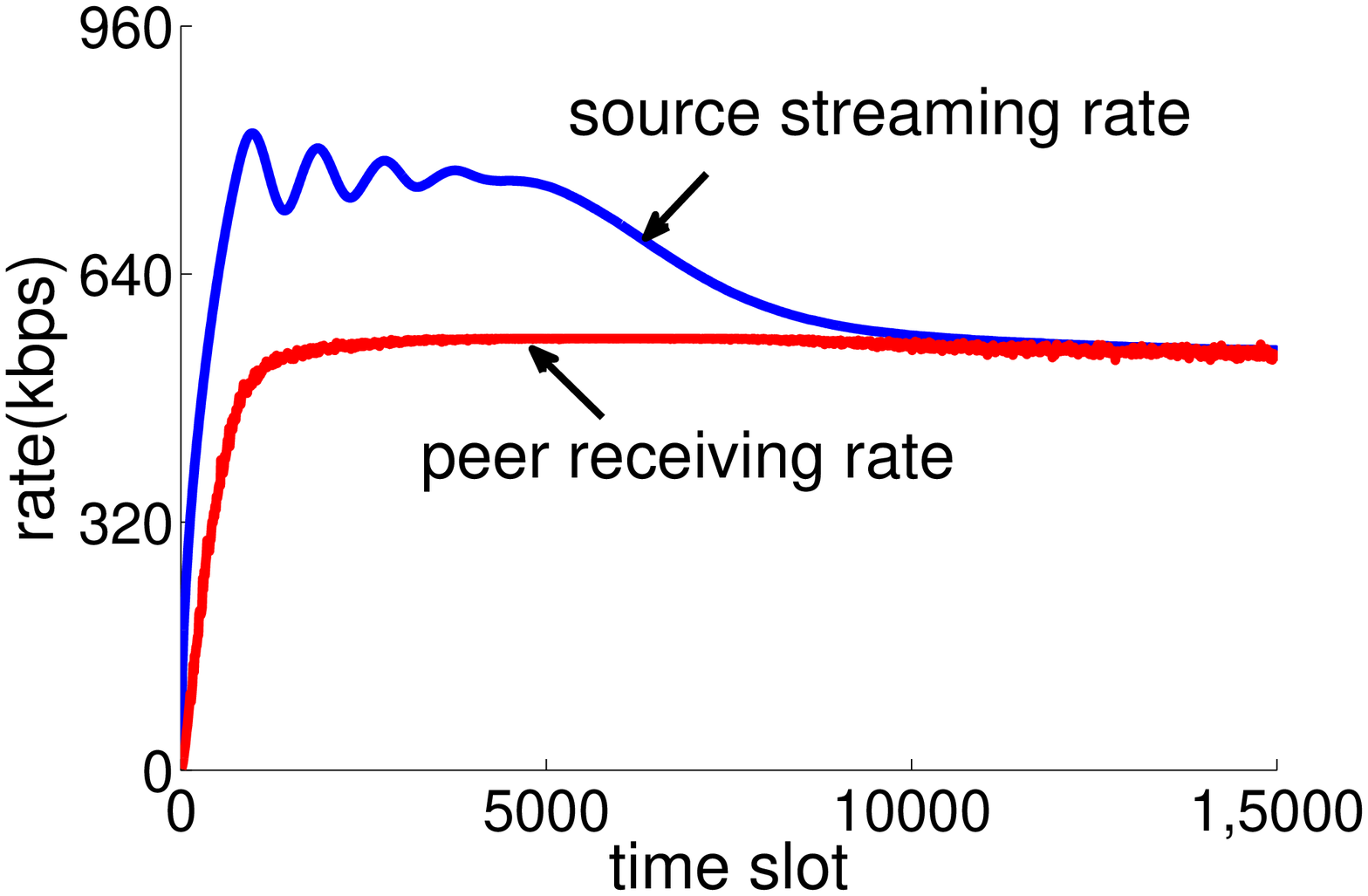}\label{fig:bp.prop.bound}}
\subfigure[]{\includegraphics[width=0.25\textwidth]{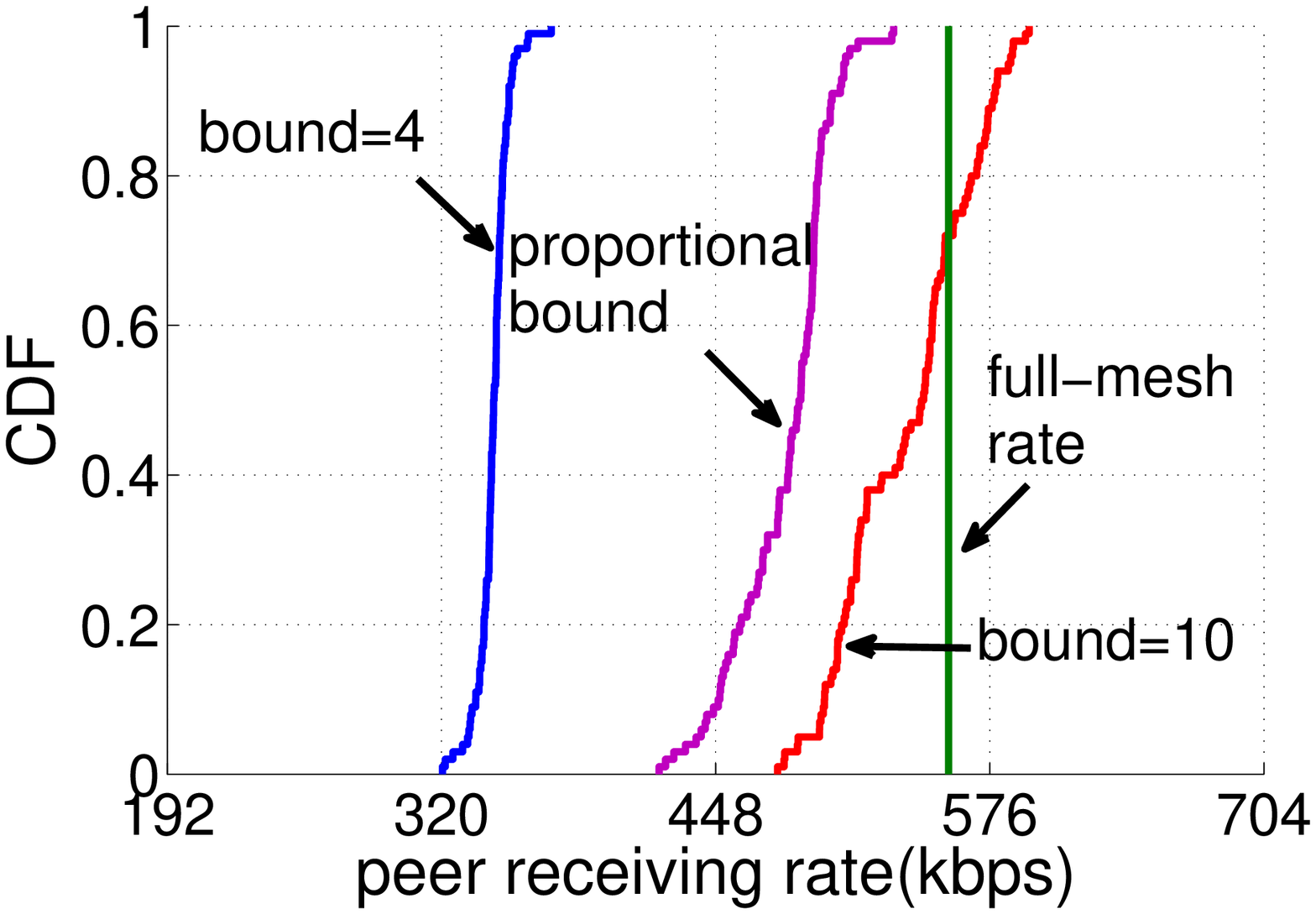}\label{fig:degree.bound.comparison}}
} \\
 \caption{Broadcasting algorithm evaluations. a) The source
broadcast rate and average peer receiving rate under Setting II when degree bound is set to $3$; b) The source broadcast rate and average peer receiving rate under Setting I when degree bound is set to $4$; c) The source broadcast rate and average peer receiving rate under Setting I when degree bound is set to 10. d) This figure shows the impact of degree bound on the peer receiving rate under Setting I. The full-mesh rate is the maximum broadcast rate when the node degrees are unbounded~\cite{all:Mutualcast:LPZ05}.}
\label{fig:bp}
\end{figure*}

\begin{figure}[hbt!]
\centering
\subfigure[]{\includegraphics[width=0.24\textwidth]{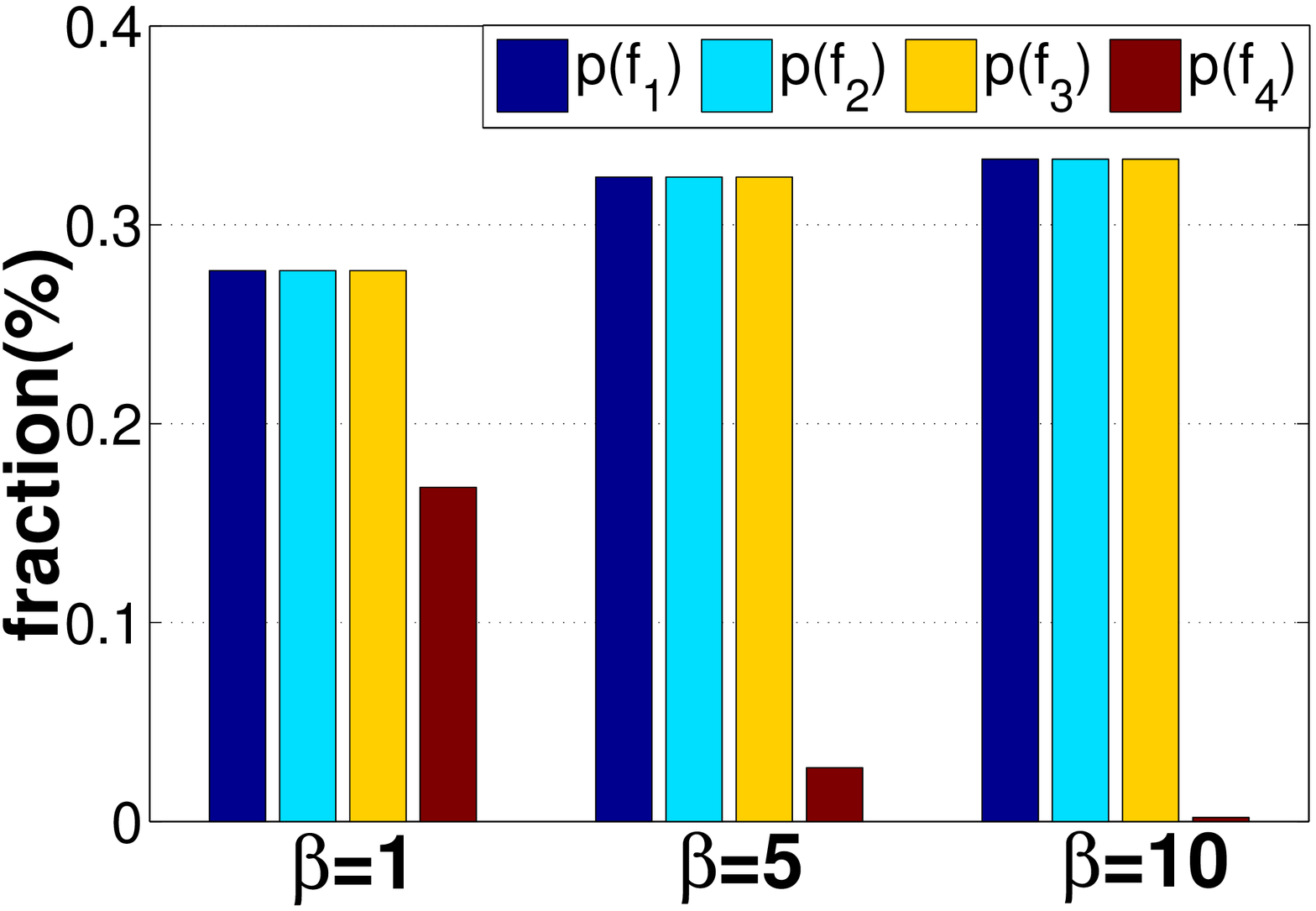}\label{fig:distrib.a}}
\subfigure[]{\includegraphics[width=0.24\textwidth]{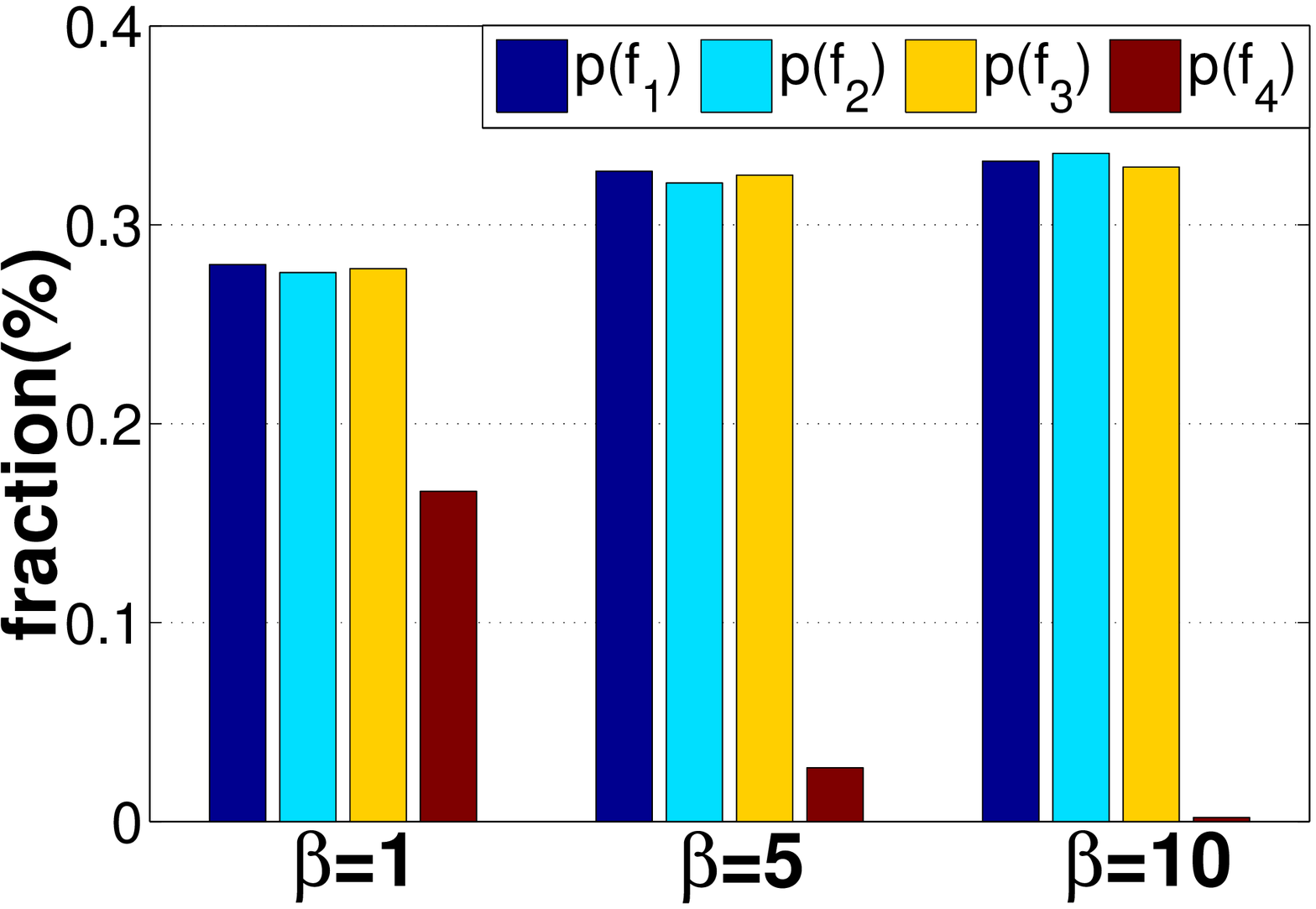}\label{fig:distrib.b}}
\caption{a) Optimal configuration distribution for different values of $\beta$ under Setting III; b) Configuration distribution obtained by our algorithm for different values of $\beta$ under Setting III.}
\end{figure}

\begin{figure*}[hbt!]
\centering \mbox{
\subfigure[]{\includegraphics[width=0.31\textwidth]{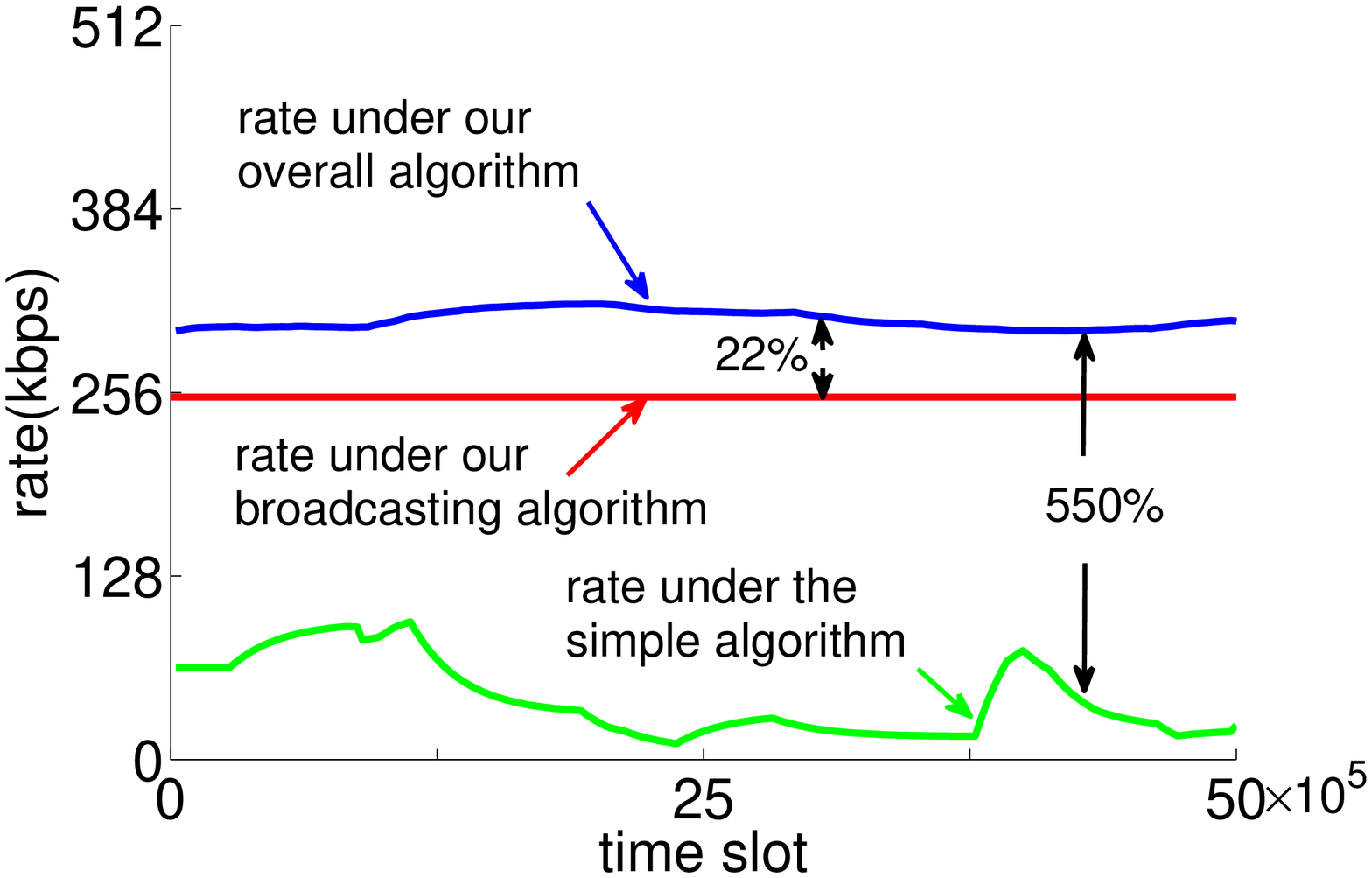}\label{fig:ma.a}}
\subfigure[]{\includegraphics[width=0.31\textwidth]{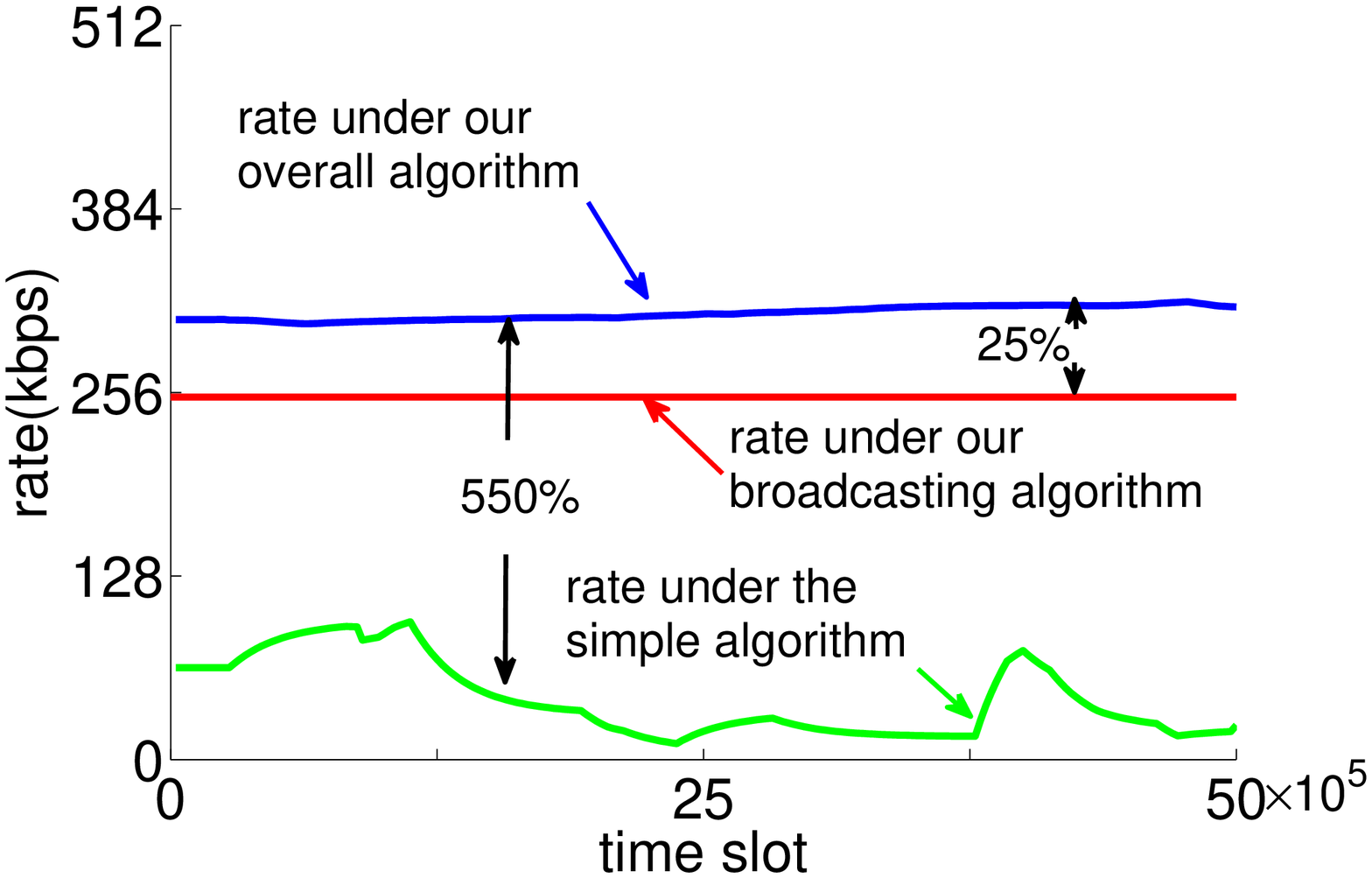}\label{fig:ma.b}}
\subfigure[]{\includegraphics[width=0.31\textwidth]{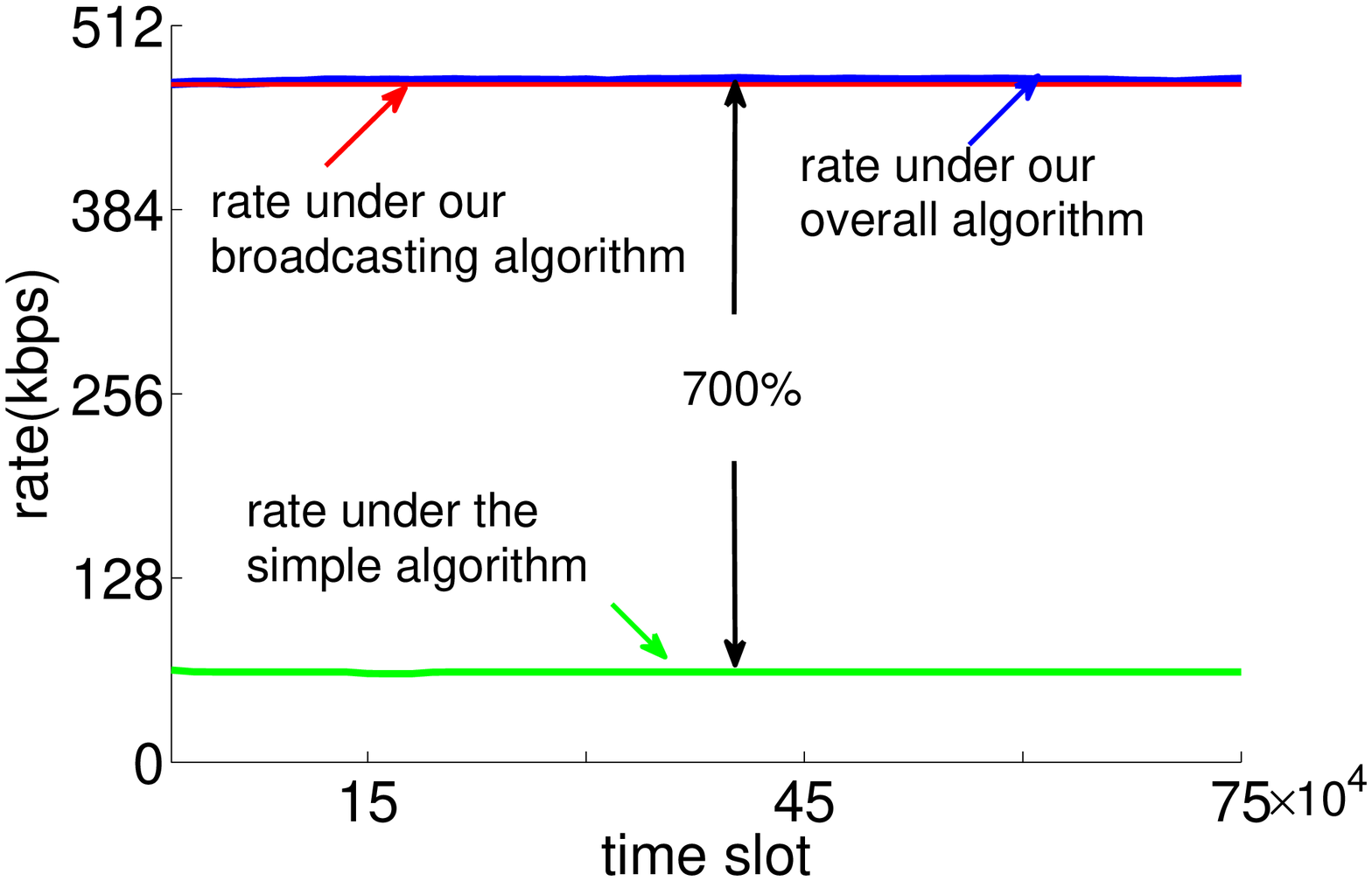}\label{fig:ma.c}}
} \\
\caption{Evaluation of our overall solution which combines the topology hopping algorithm and the broadcasting algorithm. a) The average peer receiving rate when the node degree bound is $3$ and $\beta$ is $20$; b) The average peer receiving rate when the node degree bound is $3$ and $\beta$ is $50$; c) The average peer receiving rate when peer degree bound is proportional to its upload capacity and $\beta$ is $20$. The percentage of average receiving rate improvement of our overall algorithm against our broadcasting algorithm and the simple heuristic algorithm are shown in these three figures. For example, in (a), $22\%$ means that the average receiving rate of our overall algorithm is $1.22$ times of that of our broadcasting algorithm, and $550\%$ means that the average receiving rate of our overall algorithm is $6.5$ times of that of the simple heuristic algorithm.}
\end{figure*}

\subsection{Evaluation of the Proposed Broadcasting Algorithm }

In this simulation, we evaluate our distributed broadcasting algorithm
proposed in Section~\ref{sec:streaming_rate}. We randomly choose a sub-graph that satisfies the node degree bounds constraints, and run our algorithm over it. We evaluate three aspects of the proposed algorithm: 1) does it converge to optimal broadcast rate as expected from theoretical analysis? 2) How fast does it converge?
3) How would different values of degree bounds affect the maximum
broadcast rate? The results are summarized in Fig.~\ref{fig:bp}.

From Fig.~\ref{fig:bp}(a) and Fig.~\ref{fig:bp}(b), we see that our broadcasting algorithm converges. It converges faster in the small size network as shown in Fig.~\ref{fig:bp}(a) than in the large size network as shown in Fig.~\ref{fig:bp}(b). From  Fig.~\ref{fig:bp}(d), we also see the converged rate when the node degree bond is $10$ is very close to a theoretical upper bound -- the optimal broadcast rate under no degree bounds computed according to~\cite{all:Mutualcast:LPZ05}, \cite{chen2008ump}, \cite{all:P2PStreaming:KLR.07}. This suggests that our algorithm converges to the optimal broadcast rate.

Under different degree bounds, the optimal broadcast rate varies. Fig.~\ref{fig:bp}(d) shows that the optimal broadcast rate increases when we increase the node degree bounds. We plot the CDF of peer receiving rates (after the broadcasting algorithm converges) for the case where degree bound is $4$, $10$, and proportional to the peer's upload capacity. It's seen that when the bound is $10$, the obtained rate is close to the full-mesh rate, which suggests that we do not need a large degree bound to achieve close to the full-mesh rate. The obtained rate is also close to the full-mesh rate when degree bound is proportional to the peer's upload capacity.

\subsection{Evaluation of the Overall Solution}

Our overall solution, which combines the Markov chain guided topology hopping algorithm and
the back-pressure and network coding based broadcasting algorithm, achieves the near optimal broadcast rate
under arbitrary node degree bound and over arbitrary overlay graph.
To evaluate its performance, we generate a sub-graph randomly, run our algorithms on
every node, and evaluate the achieved broadcast rate.

%

The topology hopping algorithm runs on top of the broadcasting algorithm. Under given topology, the broadcasting algorithm achieves the optimal broadcast rate. Nodes swap neighbors based on their observed receiving rate, thus changing the topology from time to time. In the simulation, we run the broadcasting algorithm long enough so that it converges before the topology transits according to the Markov chain. This way, the overall algorithm converges to the close-to-optimal broadcast rate.

In all simulations, we compare our overall algorithm with our back-pressure and network coding based broadcasting algorithm to illustrate the benefit of topology hopping, and with a simple heuristic algorithm introduced below to illustrate the benefit of our overall solution. Remind that no existing works solve the problem of streaming-rate maximization under general node degree bounds and over arbitrary topology we studied in this paper.

The simple heuristic algorithm we compare our overall algorithm against is also composed of two parts: routing-based broadcasting algorithm and random topology hopping algorithm. In routing-based broadcasting algorithm, each peer evenly allocates its upload capacity to its neighbors. Given the topology and capacity allocation, a centralized routing strategy (e.g. spanning trees based solution) is used to achieve the best broadcast rate the system can support. Similarly, the random topology hopping algorithm runs on the top of the broadcasting algorithm. Every peer maintains a timer. When the timer of one peer expires, the peer randomly drops one active neighbor which is exchanging data with it, and then selects one random candidate from its feasible neighbor list and starts to exchange data with it. By doing so, we actually allow nodes running the simple scheme to have a node degree beyond the bounds. This relaxation gives the simple scheme more degree of freedom to optimize its performance. Overall, the topology changes randomly on the top under which peers use routing to exchange streaming data.

Our first observation is that our overall scheme converges to the solution that theory predicts. We carry out simulations under Setting III.  Under this setting the optimal broadcast rate is $1$. The optimal configuration solution to problem $\mathbf{MRC-\beta}$ is calculated and shown in Fig.~\ref{fig:distrib.a} for different values of $\beta$. We run the overall scheme for this specific case and show the empirical configuration distribution in Fig.~\ref{fig:distrib.b}. Comparing the distributions in Fig.~\ref{fig:distrib.a} and Fig.~\ref{fig:distrib.b}, we can see that the distribution obtained by our overall solution is very close to the optimal one. We also calculate the achieved broadcast rate under different values of $\beta$. For $\beta=1,5$ and $10$, the broadcast rate is $0.917$, $0.987$, and $0.998$ respectively. We see that with large $\beta$, the achieved broadcast rate is close to the optimal value $1$, as predicted by our analysis in Section~\ref{sec:maf}.

Next, we evaluate our overall solution under Setting I. In Fig.~\ref{fig:ma.a} and Fig.~\ref{fig:ma.b}, the broadcast rates obtained are $305$ kbps and $312$ kbps respectively. They are about $22\%$ and $25\%$ higher respectively than the broadcast rate $250$ kbps achieved by running the broadcasting algorithm over a randomly chosen topology. This demonstrates the advantage of performing topology hopping to optimize the configuration, as compared to only randomly choosing topology.

By setting node degree bounds proportional to peers' upload capacity, nodes with higher upload capacity maintain more connections. From Fig.~\ref{fig:ma.c}, we observe that this flexibility offers a broadcast rate of $475$ kbps. Although the additional gain of topology hopping is small under the specific P2P simulation settings (e.g., node uplink capacity distribution), we remark that our topology-hopping based algorithm is theoretically guaranteed to achieve close-to-optimal streaming rate under arbitrary node degree bounds and P2P settings, while the broadcasting algorithm with random topology selection has no performance guarantee. Moreover, in practical P2P streaming systems, the node degree bounds are typically small. For example, in PPLive, the node degree bounds are 15-20 \cite{streaming_capacity.allerton09}, while the size of the system (i.e., total number of peers that are simultaneously watching the same channel) is usually hundreds of thousands. Thus, we suspect we can see substantial gain of topology hopping if our algorithm is implemented in such system with small node degree bounds, as suggested by our simulation results under small node degree bounds.

From Fig.~\ref{fig:ma.a}, Fig.~\ref{fig:ma.b} and Fig.~\ref{fig:ma.c}, we observe that the average receiving rate of our overall algorithm is about 5.5-7 times higher than that of the simple algorithm respectively. And also we can see from Fig.~\ref{fig:ma.a} and Fig.~\ref{fig:ma.b}, our algorithm can achieve smoother streaming rate than the simple algorithm because our algorithm optimizes the topology hopping and stays in the optimal topology while the simple algorithm hops among topologies randomly and arbitrarily.

\section{Discussions and Future Work}

\label{sec:conclusion}

In this paper, we propose a distributed solution to achieve a
near-optimal broadcast rate under arbitrary node degree bounds, and
over arbitrary overlay graph. Our solution is distributed and
consists of two algorithms that can be of independent interests. The
first is a distributed broadcasting algorithm that optimizes the
broadcast rate given a P2P topology. It is derived from a network
coding based problem formulation and utilizes back-pressure
arguments. It can be considered as the extension of the algorithm
in~\cite{ho2009dynamic} from link-capacity-limited underlay networks
to node-capacity-limited overlay networks. The second algorithm is a
Markov chain guided hopping algorithm that optimizes the topology,
inspired by the Markov Approximation framework introduced
in~\cite{MA:CLSC10}.

Assuming the underlying broadcasting algorithm converges instantaneously,
the topology hopping algorithm converges to the optimal configuration
distribution. When the broadcasting algorithm does not converge fast enough,
the topology hopping Markov chain transits based on inaccurate observations of the maximum broadcast rates associated with the configurations.
We show that the topology hopping algorithm still converges, but to a sub-optimal configuration distribution. We characterize an upper bound on the total variance
distance between the optimal and sub-optimal configuration distributions, as well as an upper bound on the gap between the achieved and the optimal broadcast rates. We show that both bounds decreases exponentially as the bound on inaccuracy decreases.

Using uplink bandwidth statistics of Internet hosts, our simulations
validate the effectiveness of the proposed solutions, and demonstrate
the advantage of allowing node degree bounds to scale linearly with
their upload capacities.

In the scenarios where network coding is not allowed, we can formulate the broadcasting problem in Subsection \ref{subsec:nc-formulation} as a linear program to construct a feasible node capacity allocation so that the sum of rate of all spanning
trees is maximized \cite{streaming_capacity.allerton09}, which is solvable by centralized LP algorithms. Then we can design the overall algorithm in the following way. The overall algorithm is also composed of two separate algorithms: the spanning tree based broadcasting algorithm and the Markov chain guided hopping algorithm. The topology hopping algorithm is same as the one in Section \ref{sec:maf} which runs on the top of the broadcasting algorithm and guides the topology hopping. Compared to our distributed overall algorithm when network coding is applied, this algorithm is centralized making it unsuited for use in a dynamically changing systems.

Two interesting future directions are as follows. First,
the convergence rate of our solution is determined by the mixing time
of the topology-hopping Markov chain, which can be substantial for
large P2P systems. It is thus of great interest to explore the
design of topology-hopping Markov chains that mix fast and at the
same time allows distributed implementation. Second, while our algorithms adapt well to peer dynamics, our theoretical analysis is for static scenarios. How to extend the analysis to dynamic scenarios such as those observed in practical P2P systems \cite{wang2008stable} is another interesting future direction.

\section*{Acknowledgement}

This work was partially supported by the General Research Fund grants (Project No.
411008, 411209, 411010) and an Area of Excellence Grant (Project No.
AoE/E-02/08), all established under the University Grant Committee
of the Hong Kong SAR, China. This work was also partially supported
by two gift grants from Microsoft and Cisco.

\bibliographystyle{IEEEtran}
\bibliography{IEEEabrv,ref,p2p}

\begin{thebibliography}{10}
\providecommand{\url}[1]{#1}
\csname url@samestyle\endcsname
\providecommand{\newblock}{\relax}
\providecommand{\bibinfo}[2]{#2}
\providecommand{\BIBentrySTDinterwordspacing}{\spaceskip=0pt\relax}
\providecommand{\BIBentryALTinterwordstretchfactor}{4}
\providecommand{\BIBentryALTinterwordspacing}{\spaceskip=\fontdimen2\font plus
\BIBentryALTinterwordstretchfactor\fontdimen3\font minus
  \fontdimen4\font\relax}
\providecommand{\BIBforeignlanguage}[2]{{%
\expandafter\ifx\csname l@#1\endcsname\relax
\typeout{** WARNING: IEEEtran.bst: No hyphenation pattern has been}%
\typeout{** loaded for the language `#1'. Using the pattern for}%
\typeout{** the default language instead.}%
\else
\language=\csname l@#1\endcsname
\fi
#2}}
\providecommand{\BIBdecl}{\relax}
\BIBdecl

\bibitem{all:Mutualcast:LPZ05}
J.~Li, P.~A. Chou, and C.~Zhang, ``Mutualcast: an efficient mechanism for
  content distribution in a p2p network,'' in \emph{Proc. ACM SIGCOMM Asia
  Workshop}, 2005.

\bibitem{massoulie2007rdb}
L.~Massoulie, A.~Twigg, G.~Gkantsidis, and P.~Rodriguez, ``{Randomized
  Decentralized Broadcasting Algorithms},'' in \emph{Proc. IEEE INFOCOM}, 2007.

\bibitem{all:P2PStreaming:KLR.07}
R.~Kumar, Y.~Liu, and K.~Ross, ``Stochastic fluid theory for p2p streaming
  systems,'' in \emph{Proc. {IEEE} {INFOCOM}}, 2007.

\bibitem{yicui06_optimal}
Y.~Cui, Y.~Xue, and K.~Nahrstedt, ``Optimal resource allocation in overlay
  multicast,'' \emph{IEEE Trans. Parallel and Distributed Systems}, vol.~17,
  pp. 808--823, 2006.

\bibitem{sudipta2009lcclc}
S.~Sengupta, S.~Liu, M.~Chen, M.~Chiang, J.~Li, and P.~A. Chou, ``Peer-to-peer
  streaming capacity,'' \emph{IEEE Trans. Information Theory}, vol.~57, pp.
  5072--5087, 2011.

\bibitem{castro2003shb}
M.~Castro, P.~Druschel, A.~Kermarrec, A.~Nandi, A.~Rowstron, and A.~Singh,
  ``{SplitStream: high-bandwidth multicast in cooperative environments},'' in
  \emph{Proc. ACM SOSP}, 2003.

\bibitem{padmanabhan2002ccn}
V.~Padmanabhan and K.~Sripanidkulchai, ``{The case for cooperative
  networking},'' in \emph{Proc. IPTPS}, 2002.

\bibitem{zigzag}
D.~A. Tran, K.~Hua, and T.~Do, ``{ZIGZAG}: An efficient peer-to-peer scheme for
  media streaming,'' in \emph{{Proc. IEEE INFOCOM}}, 2003.

\bibitem{magharei2009prime}
N.~Magharei and R.~Rejaie, ``{PRIME: Peer-to-peer receiver-driven mesh-based
  streaming},'' \emph{{IEEE/ACM Trans. Networking}}, vol.~17, pp. 1052--1065,
  2009.

\bibitem{streaming_capacity.icdcs10}
S.~Liu, M.~Chen, S.~Sengupta, M.~Chiang, J.~Li, and P.~A. Chou, ``Peer-to-peer
  streaming capacity under node degree bound,'' in \emph{Proc. IEEE ICDCS},
  2010.

\bibitem{fenglili}
C.~Feng, B.~Li, and B.~Li, ``Understanding the performance gap between
  pull-based mesh streaming protocols and fundamental limits,'' in \emph{Proc.
  IEEE INFOCOM}, 2009.

\bibitem{AbeKirCig}
L.~Abeni, C.~Kiraly, and R.~L. Cigno, ``On the optimal scheduling of streaming
  applications in unstructured meshes,'' in \emph{Proc. IFIP Networking}, 2009.

\bibitem{wang2007r}
M.~Wang and B.~Li, ``{R2: Random Push with Random Network Coding in Live
  Peer-to-Peer Streaming},'' \emph{IEEE Journal on Selected Areas in
  Communications}, vol.~25, p. 1655, 2007.

\bibitem{zhang2005coolstreaming}
X.~Zhang, J.~Liu, B.~Li, and T.~Yum, ``{CoolStreaming/DONet: A data-driven
  overlay network for efficient live media streaming},'' in \emph{Proc. IEEE
  INFOCOM}, 2005.

\bibitem{streaming_capacity.allerton09}
M.~Chen, M.~Chiang, P.~A. Chou, J.~Li, S.~Liu, and S.~Sengupta, ``Peer-to-peer
  streaming capacity: Survey and recent results,'' in \emph{Proc. Allerton
  Conference}, 2009.

\bibitem{all:NetCodP2P:CYHF.06}
D.~M. Chiu, R.~W. Yeung, J.~Huang, and B.~Fan, ``Can network coding help in p2p
  networks?'' in \emph{Proc. IEEE NetCod}, 2006.

\bibitem{chen2008ump}
M.~Chen, M.~Ponec, S.~Sengupta, J.~Li, and P.~Chou, ``{Utility maximization in
  peer-to-peer systems},'' in \emph{Proc. ACM SIGMETRICS}, 2008.

\bibitem{all:pplive}
\BIBentryALTinterwordspacing
P{P}{L}ive. [Online]. Available: \url{http://www.pplive.com}
\BIBentrySTDinterwordspacing

\bibitem{all:uusee}
\BIBentryALTinterwordspacing
U{U}{S}ee. [Online]. Available: \url{http://www.uusee.com}
\BIBentrySTDinterwordspacing

\bibitem{MA:CLSC10}
M.~Chen, S.~C. Liew, Z.~Shao, and C.~Kai, ``Markov approximation for
  combinatorial network optimization,'' in \emph{Proc. IEEE INFOCOM}, 2010.

\bibitem{all:NetCod:ACLY00}
R.~Ahlswede, N.~Cai, S.-Y.~R. Li, and R.~W. Yeung, ``Network information
  flow,'' \emph{IEEE Trans. Information Theory}, pp. 1204--1216, 2000.

\bibitem{Tassiulas92backpressure}
L.~Tassiulas and A.~Ephremides, ``{Stability properties of constrained queueing
  systems and scheduling policies for maximum throughput in multihop radio
  networks},'' \emph{IEEE Trans. Automatic Control}, vol.~37, pp. 1936--1948,
  1992.

\bibitem{all:Backpressure:MAW96}
N.~McKeown, V.~Anantharam, and J.~Walrand, ``Achieving 100\% throughput in an
  input-queued switch,'' in \emph{Proc. IEEE INFOCOM}, 1996.

\bibitem{all:Backpressure:ESP05}
A.~Eryilmaz, R.~Srikant, and J.~Perkins, ``Stable scheduling policies for
  fading wireless channels,'' \emph{IEEE/ACM Trans. on Networking}, vol.~13,
  pp. 411--424, 2005.

\bibitem{all:Backpressure:NMR05}
M.~Neely, E.~Modiano, and C.~Rohrs, ``Dynamic power allocation and routing for
  time-varying wireless networks,'' \emph{IEEE Journal on Selected Areas in
  Communications}, vol.~23, pp. 89--103, 2005.

\bibitem{liu2008opportunities}
J.~Liu, S.~Rao, B.~Li, and H.~Zhang, ``Opportunities and challenges of
  peer-to-peer internet video broadcast,'' in \emph{Proc. IEEE}, 2008.

\bibitem{liu-uusee}
Z.~Liu, C.~Wu, B.~Li, and S.~Zhao, ``Uusee: Large-scale operational on-demand
  streaming with random network coding,'' in \emph{Proc. {IEEE} {INFOCOM}},
  2010.

\bibitem{hei2007msl}
X.~Hei, C.~Liang, J.~Liang, Y.~Liu, and K.~Ross, ``{A Measurement Study of a
  Large-Scale P2P IPTV System},'' \emph{IEEE Trans. Multimedia}, vol.~9, pp.
  1672--1687, 2007.

\bibitem{chou2003practical}
P.~Chou, Y.~Wu, and K.~Jain, ``{Practical network coding},'' in \emph{Proc.
  Allerton Conference}, 2003.

\bibitem{ho2009dynamic}
T.~Ho and H.~Viswanathan, ``{Dynamic algorithms for multicast with
  intra-session network coding},'' \emph{IEEE Trans. Information Theory},
  vol.~55, pp. 797--815, 2005.

\bibitem{Boyd04}
S.~Boyd and L.~Vandenberghe, \emph{{Convex optimization}}.\hskip 1em plus 0.5em
  minus 0.4em\relax Cambridge university press, 2004.

\bibitem{all:article:HMKKESL04}
T.~Ho, M.~M{\'e}dard, R.~Koetter, D.~Karger, M.~Effros, J.~Shi, and B.~Leong,
  ``A random linear network coding approach to multicast,'' \emph{IEEE Trans.
  Information Theory}, vol.~52, 2004.

\bibitem{all:RandomNC:SET03}
P.~Sanders, S.~Egner, and L.~Tolhuizen, ``Polynomial time algorithms for
  network information flow,'' in \emph{Proc. ACM Symposium on Parallel
  Algorithms and Architectures}, 2003.

\bibitem{all:mathematics.cong.ctrl}
R.~Srikant, \emph{The Mathematics of Internet Congestion Control}.\hskip 1em
  plus 0.5em minus 0.4em\relax Birkh{\"{a}}user, 2004.

\bibitem{georgiadis2006resource}
L.~Georgiadis, M.~Neely, and L.~Tassiulas, \emph{{Resource allocation and cross
  layer control in wireless networks}}.\hskip 1em plus 0.5em minus 0.4em\relax
  Now Pub, 2006.

\bibitem{Diaconis91}
P.~Diaconis and D.~Stroock, ``{Geometric bounds for eigenvalues of Markov
  chains},'' \emph{The Annals of Applied Probability}, pp. 36--61, 1991.

\bibitem{vodprofitable}
C.~Huang, J.~Li, and K.~W. Ross, ``Can internet video-on-demand be
  profitable?'' in \emph{Proc. ACM SIGCOMM}, 2007.

\bibitem{wang2008stable}
F.~Wang, J.~Liu, and Y.~Xiong, ``Stable peers: Existence, importance, and
  application in peer-to-peer live video streaming,'' in \emph{Proc. IEEE
  INFOCOM}, 2008.

\bibitem{Kelly79}
F.~Kelly, \emph{{Reversibility and stochastic networks}}.\hskip 1em plus 0.5em
  minus 0.4em\relax Wiley,Chichester, 1979.

\end{thebibliography}

\appendices
\section*{Appendix}

\subsection{Proof of Theorem \ref{thm:converge}} \label{sec:proof_converge}

We use the following Lyapunov function

\begin{align*}
V(z,\boldsymbol{\lambda},\boldsymbol{\theta})= & \frac{1}{2\alpha}(z-z^{*})^{2}+\frac{1}{2}\sum_{v\in V}\sum_{d\in R}\frac{1}{k_{v,d}}(\lambda_{v,d}-\lambda_{v,d}^{*})^{2},\end{align*}
where $z^{*},\:\boldsymbol{\lambda}^{*}$ are the saddle points
of \eqref{eq:partial.lag}.

By differentiating the Lyapunov function with respect to time we get

\begin{align}
 & \dot{V}(z,\boldsymbol{\lambda})\nonumber \\
= & (z-z^{*})\left[U^{'}(z)-\sum_{d\in R}\lambda_{s,d}\right]_{z}^{+}\nonumber \\
 & +\sum_{v\in V}\sum_{d\in R}(\lambda_{v,d}-\lambda_{v,d}^{*})\left[\sum_{u\in in(v)}f_{uv}^{d}+z\mathbf{1}_{v=s}-\sum_{u\in out(v)}f_{vu}^{d}\right]_{\lambda_{v,d}}^{+}\nonumber \\
\leq & (z-z^{*})\left[U^{'}(z)-\sum_{d\in R}\lambda_{s,d}\right]\nonumber \\
 & +\sum_{v\in V}\sum_{d\in R}(\lambda_{v,d}-\lambda_{v,d}^{*})\left[\sum_{u\in in(v)}f_{uv}^{d}+z\mathbf{1}_{v=s}-\sum_{u\in out(v)}f_{vu}^{d}\right].\label{eq:diff.inequality1}\end{align}

KKT conditions for $z^{*},\boldsymbol{\lambda}^{*}$ are shown as follows

\begin{equation}
U^{'}(z^{*})-\sum_{d\in R}\lambda_{s,d}^{*}=0\label{eq:kkt1}\end{equation}
and
\begin{equation}
\lambda_{v,d}^{*}\left[\sum_{u\in in(v)}(f_{uv}^{d})^{*}+z^{*}\mathbf{1}_{v=s}-\sum_{u\in out(v)}(f_{vu}^{d})^{*}\right]=0,\: v\in V,d\in R\label{eq:kkt2},\end{equation}
where $\boldsymbol{f}^{*}$is the optimal solution of $\mathbf{MP}$.

From equation \eqref{eq:kkt1}, we obtain

\begin{equation}
U^{'}(z^{*})=\sum_{d\in R}\lambda_{s,d}^{*}.\label{eq:eq1}\end{equation}
By using the above equation\eqref{eq:eq1}, we can transform the
terms in the inequality \eqref{eq:diff.inequality1} as follows

\begin{align}
 & (z-z^{*})\left[U^{'}(z)-\sum_{d\in R}\lambda_{s,d}\right]\nonumber \\
= & (z-z^{*})\left(U^{'}(z)-U^{'}(z^{*})\right)+(z-z^{*})\left[\sum_{d\in R}\lambda_{s,d}^{*}-\sum_{d\in R}\lambda_{s,d}\right]\label{eq:eq2}\end{align}
and
\begin{align}
 & \sum_{v\in V}\sum_{d\in R}(\lambda_{v,d}-\lambda_{v,d}^{*})\left[\sum_{u\in in(v)}f_{uv}^{d}+z\mathbf{1}_{v=s}-\sum_{u\in out(v)}f_{vu}^{d}\right]\nonumber \\
= & \sum_{v\in V}\sum_{d\in R}(\lambda_{v,d}-\lambda_{v,d}^{*})\left[\sum_{u\in in(v)}f_{uv}^{d}+z^{*}\mathbf{1}_{v=s}-\sum_{u\in out(v)}f_{vu}^{d}\right]\nonumber \\
 & +(z-z^{*})\sum_{d\in R}(\lambda_{s,d}-\lambda_{s,d}^{*}).\label{eq:eq3}\end{align}

We use the above two equations \eqref{eq:eq2} and \eqref{eq:eq3} to
substitute the corresponding terms in the inequality
\eqref{eq:diff.inequality1} and then get

\begin{align}
 & \dot{V}(z,\boldsymbol{\lambda})\nonumber \\
\leq & (z-z^{*})\left(U^{'}(z)-U^{'}(z^{*})\right)\nonumber \\
 & +\sum_{v\in V}\sum_{d\in R}(\lambda_{v,d}-\lambda_{v,d}^{*})\left[\sum_{u\in in(v)}f_{uv}^{d}+z^{*}\mathbf{1}_{v=s}-\sum_{u\in out(v)}f_{vu}^{d}\right]\nonumber \\
= & (z-z^{*})\left(U^{'}(z)-U^{'}(z^{*})\right)\label{eq:bound1}\\
 & +\sum_{v\in V}\sum_{d\in R}\lambda_{v,d}\left[\sum_{u\in in(v)}f_{uv}^{d}+z^{*}\mathbf{1}_{v=s}-\sum_{u\in out(v)}f_{vu}^{d}\right]\label{eq:bound2}\\
 & +\sum_{v\in V}\sum_{d\in R}\lambda_{v,d}^{*}\left[\sum_{u\in out(v)}f_{vu}^{d}-\sum_{u\in in(v)}f_{uv}^{d}-z^{*}\mathbf{1}_{v=s}\right].\label{eq:bound3}\end{align}

Next we check the value of \eqref{eq:bound1}, \eqref{eq:bound2},
\eqref{eq:bound3} respectively. First, the strict concavity of $U(\cdot)$
implies

\begin{equation}
(z-z^{*})\left(U^{'}(z)-U^{'}(z^{*})\right)\leq0.\end{equation}

Since $z^{*},\boldsymbol{\lambda}^{*}$ are optimal solutions, they
should satisfy the constraints of the problem $\mathbf{MP}$. So we
have

\[
\sum_{u\in in(v)}(f_{uv}^{d})^{*}+z^{*}\mathbf{1}_{v=s}\leq\sum_{u\in out(v)}(f_{vu}^{d})^{*}\: v\in V,d\in R.\]
Therefore,

\begin{align*}
 & \sum_{v\in V}\sum_{d\in R}\lambda_{v,d}\left[\sum_{u\in in(v)}f_{uv}^{d}+z^{*}\mathbf{1}_{v=s}-\sum_{u\in out(v)}f_{vu}^{d}\right]\\
\leq & \sum_{v\in V}\sum_{d\in R}\lambda_{v,d}\left[\sum_{u\in in(v)}f_{uv}^{d}-\sum_{u\in out(v)}f_{vu}^{d}+\sum_{u\in out(v)}(f_{vu}^{d})^{*}-\sum_{u\in in(v)}(f_{uv}^{d})^{*}\right]\\
= & \sum_{v\in V}\sum_{d\in R}\lambda_{v,d}\left[\sum_{u\in in(v)}f_{uv}^{d}-\sum_{u\in out(v)}f_{vu}^{d}\right]\\
 & -\sum_{v\in V}\sum_{d\in R}\lambda_{v,d}\left[\sum_{u\in in(v)}(f_{uv}^{d})^{*}-\sum_{u\in out(v)}(f_{vu}^{d})^{*}\right].\end{align*}
Note that $\boldsymbol{f}$ is the solution of the following problem

\begin{align*}
\max & \sum_{v\in V}\sum_{d\in R}\lambda_{v,d}\left[\sum_{u\in out(v)}f_{vu}^{d}-\sum_{u\in in(v)}f_{uv}^{d}\right]\\
\mbox{s.t.} & \eqref{eq:MPC2}-\eqref{eq:MPC3},\end{align*}
which is equivalent to $\mathbf{SSP}$. Since $\boldsymbol{f}^{*}$is
also feasible, for \eqref{eq:bound2} we have

\begin{align*}
 & \sum_{v\in V}\sum_{d\in R}\lambda_{v,d}\left[\sum_{u\in in(v)}f_{uv}^{d}+z^{*}\mathbf{1}_{v=s}-\sum_{u\in out(v)}f_{vu}^{d}\right]\\
\leq & \sum_{v\in V}\sum_{d\in R}\lambda_{v,d}\left[\sum_{u\in in(v)}f_{uv}^{d}-\sum_{u\in out(v)}f_{vu}^{d}\right]\\
 & -\sum_{v\in V}\sum_{d\in R}\lambda_{v,d}\left[\sum_{u\in in(v)}(f_{uv}^{d})^{*}-\sum_{u\in out(v)}(f_{vu}^{d})^{*}\right]\\
\leq & 0.\end{align*}

Now we focus on the term\eqref{eq:bound3}. According to \eqref{eq:kkt2},
the following equality holds.
\begin{align*}
 & \sum_{v\in V}\sum_{d\in R}\lambda_{v,d}^{*}\left[\sum_{u\in out(v)}f_{vu}^{d}-\sum_{u\in in(v)}f_{uv}^{d}-z^{*}\mathbf{1}_{v=s}\right]\\
= & \sum_{v\in V}\sum_{d\in R}\lambda_{v,d}^{*}\left[\sum_{u\in out(v)}f_{uv}^{d}-\sum_{u\in in(v)}f_{vu}^{d}+\sum_{u\in in(v)}(f_{vu}^{d})^{*}-\sum_{u\in out(v)}(f_{uv}^{d})^{*}\right]\\
= & \sum_{v\in V}\sum_{d\in R}\lambda_{v,d}^{*}\left[\sum_{u\in out(v)}f_{uv}^{d}-\sum_{u\in in(v)}f_{vu}^{d}\right]\\
 & -\sum_{v\in V}\sum_{d\in R}\lambda_{v,d}^{*}\left[\sum_{u\in out(v)}(f_{uv}^{d})^{*}-\sum_{u\in in(v)}(f_{vu}^{d})^{*}\right].\end{align*}
Note that $\boldsymbol{f}^{*}$ is the solution of the following problem

\begin{align*}
\max & \sum_{v\in V}\sum_{d\in R}\lambda_{v,d}^{*}\left[\sum_{u\in out(v)}f_{vu}^{d}-\sum_{u\in in(v)}f_{uv}^{d}\right]\\
\mbox{s.t.} & \eqref{eq:MPC2}-\eqref{eq:MPC3}.\end{align*}
So, \begin{align*}
 & \sum_{v\in V}\sum_{d\in R}\lambda_{v,d}^{*}\left[\sum_{u\in out(v)}f_{vu}^{d}-\sum_{u\in in(v)}f_{uv}^{d}-z^{*}\mathbf{1}_{v=s}\right]\\
= & \sum_{v\in V}\sum_{d\in R}\lambda_{v,d}^{*}\left[\sum_{u\in out(v)}f_{uv}^{d}-\sum_{u\in in(v)}f_{vu}^{d}\right]\\
 & -\sum_{v\in V}\sum_{d\in R}\lambda_{v,d}^{*}\left[\sum_{u\in out(v)}(f_{uv}^{d})^{*}-\sum_{u\in in(v)}(f_{vu}^{d})^{*}\right]\\
\leq & 0.\end{align*}

Overall, we get

\begin{align*}
 & \dot{V}(z,\boldsymbol{\lambda})\\
\leq & (z-z^{*})\left(U^{'}(z)-U^{'}(z^{*})\right)\\
 & +\sum_{v\in V}\sum_{d\in R}\lambda_{v,d}\left[\sum_{u\in in(v)}f_{uv}^{d}+z^{*}\mathbf{1}_{v=s}-\sum_{u\in out(v)}f_{vu}^{d}\right]\\
 & +\sum_{v\in V}\sum_{d\in R}\lambda_{v,d}^{*}\left[\sum_{u\in out(v)}f_{vu}^{d}-\sum_{u\in in(v)}f_{uv}^{d}-z^{*}\mathbf{1}_{v=s}\right]\\
\leq & 0.\end{align*}

Let $\mathcal{E}\triangleq\{(z,\boldsymbol{\lambda})|\,\dot{V}(z,\boldsymbol{\lambda})=0\}$
and $\mathcal{G}\triangleq\{(z,\boldsymbol{\lambda})|\,\eqref{eq:bound1}=0,\eqref{eq:bound2}=0,\eqref{eq:bound3}=0\}$.
Since $\dot{V}(z,\boldsymbol{\lambda})\leq\eqref{eq:bound1}+\eqref{eq:bound2}+\eqref{eq:bound3}$
and $\eqref{eq:bound1}\leq0,\eqref{eq:bound2}\leq0,\eqref{eq:bound3}\leq0,$
we have $\mathcal{E}\subset\mathcal{G}$. Let $\mathcal{M}$ be the
largest invariant set in $\mathcal{E}$. By LaSalle\textquoteright{}s
invariance principle $(z(t),\boldsymbol{\lambda}(t))$ converges to
the set $\mathcal{M}$ as $t\rightarrow\infty$. Since $\mathcal{M}\subset\mathcal{E}\subset\mathcal{G}$,
as $t\rightarrow\infty$ $(z(t),\boldsymbol{\lambda}(t))$ satisfies

\begin{equation}
z(t)=z^{*}\label{eq:asym.eq}\end{equation}
and
\begin{small}
\begin{equation}
\sum_{v\in V}\sum_{d\in R}\left(\lambda_{v,d}(t)-\lambda_{v,d}^{*}(t)\right)\left[\sum_{u\in in(v)}f_{uv}^{d}(t)+z^{*}\mathbf{1}_{v=s}-\sum_{u\in out(v)}f_{vu}^{d}(t)\right]=0.\end{equation}
\end{small}
Further, in $\mathcal{M}$, $\sum_{d\in
R}\lambda_{s,d}(t)=U^{'}(z^{*})$. To see this, if this is not
satisfied, then by \eqref{eq:PD} we can see $z(t)$ will not stay in
$z^{*}$, which is contradicted with \eqref{eq:asym.eq}. This
concludes the proof.

\subsection{Proof of Proposition \ref{implement_mc}}
\label{sec:proof_mc}
By two conditions for state space structure of Markov chain, we know
that all configurations can reach each other within a finite number
of transitions, thus the constructed Markov chain is irreducible.
Further, it is a finite state ergodic Markov chain with a unique
stationary distribution. We now show that the stationary
distribution of the constructed Markov chain is indeed
\eqref{dist_no_error}.

Now we verify that under the implementation, the state transition
rate from $f$ to $f^{'}$ satisfies (\ref{eq:transition rate}).

In our Markov chain design, we only allow direct transitions between two configurations
if such transitions correspond to a single node adding a new neighbor
or removing a neighbor, i.e., $|\mathcal{N}_{f}\backslash\mathcal{N}_{f'}|=1$ or $|\mathcal{N}_{f'}\backslash\mathcal{N}_{f}|=1$. We consider these two scenarios
separately in the following.

Let $f\rightarrow f^{'}$ denote the event that when the timer expires
the process will enter state $f^{'}$ after leaving the current state
$f$. The probability of this event is denoted by $\Pr(f\rightarrow f^{'})$.

When $|\mathcal{N}_{f}\backslash\mathcal{N}_{f'}|=1$,
assuming $\mathcal{N}_{f}\backslash\mathcal{N}_{f'}=(v,u)$,
the event $f\rightarrow f^{'}$ can be divided into two disjoint events: the
event that node $v$'s timer expires, then node $v$ selects node
$u$ to remove and remove it from its in-use neighbor set and the
event that node $u$'s timer expires, then node $u$ selects node
$v$ to remove and remove it from its in-use neighbor set. Denote
these two events by $f\,\underrightarrow{v-u}\, f^{'}$ and $f\,\underrightarrow{u-v}\, f^{'}$
. Let $v-u$ be the event that node $v$ selects node $u$ and removes
it from its in-use neighbor set and $u-v$ be the event that node
$u$ selects node $v$ and removes it from its in-use neighbor set.
Now we calculate the probability of $f\,\underrightarrow{v-u}\, f^{'}$
and $f\,\underrightarrow{u-v}\, f^{'}$ respectively.

\begin{align}
 & \Pr(f\,\underrightarrow{v-u}\, f^{'})\nonumber \\
= & \Pr(v-u|v\mbox{'s timer expires})\Pr(v\mbox{'s timer expires})\nonumber \\
= & \frac{|N_{v,f}|}{|N_{v}|}\cdot\frac{1}{|N_{v,f}|}\cdot\frac{\exp(\beta x_{f^{'}})}{\exp(\beta x_{f})+\exp(\beta x_{f^{'}})}\cdot\frac{\frac{|N_{v}|}{2\exp(\tau)}}{\sum_{w\in V}\frac{|N_{w}|}{2\exp(\tau)}}\nonumber \\
= & \frac{1}{\sum_{v\in V}|N_{v}|}\cdot\frac{\exp(\beta x_{f^{'}})}{\exp(\beta x_{f})+\exp(\beta x_{f^{'}})}
\end{align}
and
\begin{align}
 & \Pr(f\,\underrightarrow{u-v}\, f^{'})\nonumber \\
= & \Pr(u-v|u\mbox{'s timer expires})\Pr(u\mbox{'s timer expires})\nonumber \\
= & \frac{|N_{u,f}|}{|N_{u}|}\cdot\frac{1}{|N_{u,f}|}\cdot\frac{\exp(\beta x_{f^{'}})}{\exp(\beta x_{f})+\exp(\beta x_{f^{'}})}\cdot\frac{\frac{|N_{u}|}{2\exp(\tau)}}{\sum_{w\in V}\frac{|N_{w}|}{2\exp(\tau)}}\nonumber \\
= & \frac{1}{\sum_{v\in V}|N_{v}|}\cdot\frac{\exp(\beta x_{f^{'}})}{\exp(\beta x_{f})+\exp(\beta x_{f^{'}})}.
\end{align}
Therefore, we have
\begin{align}
 & \Pr(f\rightarrow f^{'})\nonumber \\
= & \Pr(f\,\underrightarrow{v-u}\, f^{'})+\Pr(f\,\underrightarrow{u-v}\, f^{'})\nonumber \\
= & \frac{2}{\sum_{v\in V}|N_{v}|}\cdot\frac{\exp(\beta x_{f^{'}})}{\exp(\beta x_{f})+\exp(\beta x_{f^{'}})}.
\end{align}

When $|\mathcal{N}_{f'}\backslash\mathcal{N}_{f}|=1$,
assuming $\mathcal{N}_{f'}\backslash\mathcal{N}_{f}=(v,u)$,
similarly we divide $f\rightarrow f^{'}$ into two disjoint events $f\,\underrightarrow{v+u}\, f^{'}$
and $f\,\underrightarrow{u+v}\, f^{'}$. $f\,\underrightarrow{v+u}\, f^{'}$
denotes the event that node $v$'s timer expires, then node $v$ selects
node $u$ to add and add it in its in-use neighbor set. $f\,\underrightarrow{u+v}\, f^{'}$
denotes the event that node $u$'s timer expires, then node $u$ selects
node $v$ to add and add it in its in-use neighbor set. Let $v+u$
be the event that node $v$ selects node $u$ and adds it as one in-use
neighbor and $u+v$ be the event that node $u$ selects node $v$
and adds it as one in-use neighbor. Then we have

\begin{align}
 & \Pr(f\,\underrightarrow{v+u}\, f^{'})\nonumber \\
= & \Pr(v+u|v\mbox{'s timer expires})\Pr(v\mbox{'s timer expires})\nonumber \\
= & \frac{|N_{v}|-|N_{v,f}|}{|N_{v}|}\cdot\frac{1}{|N_{v}|-|N_{v,f}|}\cdot\frac{\exp(\beta x_{f^{'}})}{\exp(\beta x_{f})+\exp(\beta x_{f^{'}})}\cdot\frac{\frac{|N_{v}|}{2\exp(\tau)}}{\sum_{w\in V}\frac{|N_{w}|}{2\exp(\tau)}}\nonumber \\
= & \frac{1}{\sum_{v\in V}|N_{v}|}\cdot\frac{\exp(\beta x_{f^{'}})}{\exp(\beta x_{f})+\exp(\beta x_{f^{'}})}
\end{align}
and
\begin{align}
 & \Pr(f\,\underrightarrow{u+v}\, f^{'})\nonumber \\
= & \Pr(u+v|u\mbox{'s timer expires})\Pr(u\mbox{'s timer expires})\nonumber \\
= & \frac{|N_{u}|-|N_{u,f}|}{|N_{u}|}\cdot\frac{1}{|N_{u}|-|N_{u,f}|}\cdot\frac{\exp(\beta x_{f^{'}})}{\exp(\beta x_{f})+\exp(\beta x_{f^{'}})}\cdot\frac{\frac{|N_{u}|}{2\exp(\tau)}}{\sum_{w\in V}\frac{|N_{w}|}{2\exp(\tau)}}\nonumber \\
= & \frac{1}{\sum_{v\in V}|N_{v}|}\cdot\frac{\exp(\beta x_{f^{'}})}{\exp(\beta x_{f})+\exp(\beta x_{f^{'}})}.
\end{align}
Therefore, we have
\begin{align}
 & \Pr(f\rightarrow f^{'})\nonumber \\
= & \Pr(f\,\underrightarrow{v+u}\, f^{'})+\Pr(f\,\underrightarrow{u+v}\, f^{'})\nonumber \\
= & \frac{2}{\sum_{v\in V}|N_{v}|}\cdot\frac{\exp(\beta x_{f^{'}})}{\exp(\beta x_{f})+\exp(\beta x_{f^{'}})}.
\end{align}

In our implementation, under configuration $f$, peer $v$ counts down
with rate $\frac{|N_v|}{2}\exp^{-1}(\tau)$.
Therefore, the rate of leaving the state $f$ is
$\sum_{v\in{V}}\frac{|N_v|}{2}\exp^{-1}(\tau)$.
With the probability $\Pr(f\rightarrow f^{'})$, the process jumps to state $f^{'}$
when leaving state $f$. So, the transition rate from state $f$ to
$f^{'}$ is
\begin{align}
 & q_{f,f'}=\frac{2}{\sum_{v\in V}|N_v|} \cdot\frac{\exp(\beta\cdot x_{f^{'}})}{\exp(\beta\cdot x_{f})+\exp(\beta\cdot x_{f^{'}})}\nonumber \\
 & \times\sum_{v\in V}\frac{|N_v|}{2}\exp^{-1}(\tau)\nonumber \\
 & =\exp^{-1}(\tau)\frac{\exp(\beta\cdot x_{f^{'}})}{\exp(\beta\cdot x_{f})+\exp(\beta\cdot x_{f^{'}})}.
\end{align}

With \eqref{dist_no_error}, we see that
$p_{f}^{*}(\boldsymbol{x})\cdot
q_{f,f'}=p_{f'}^{*}(\boldsymbol{x})\cdot q_{f',f}, \forall f,f' \in
\mathcal{F}$, i.e., the detailed balance equations hold. Thus the
constructed Markov chain is time-reversible and its stationary
distribution is indeed \eqref{dist_no_error} according to Theorem
1.3 and Theorem 1.14 in \cite{Kelly79}.

\subsection{Proof of Theorem
\ref{error_impact}}\label{sec:error_impact}

We denote $M$ as the original topology hopping Markov chain with
exact broadcast rates, and $M'$ as the corresponding extended Markov
chain with inaccurately observed broadcast rates. For the
convenience of expression, for all $f \in \mathcal{F},
j\in\{-n_{f},\ldots,n_{f}\}$, we use $f_j$ to represent the state
$(f, x_f+\frac{j}{n_f}\Delta_f)$ in the extended Markov chain $M'$,
and $\eta_{f_j}$ to represent distribution of inaccurate observed
rates $\eta_{j,f}$.

Therefore, given direct transitions between configuration $f$ and
$f'$ in the original topology hopping Markov chain $M$, there are
direct transitions between states $f_j$ and $f'_k$ ($\forall
j\in\{-n_{f},\ldots,n_{f}\},k\in\{-n_{f'},\ldots,n_{f'}\}$) in the
extended Markov chain $M'$. Following \eqref{tran_rate_mc1} and
\eqref{tran_rate_mc2}, we have the corresponding transition rates
\begin{align}
& q_{f_j,f'_k}=\frac{\eta_{f'_k}}{\exp(\tau)}\cdot
\frac{\exp(\beta(x_{f'}+\frac{k}{n_{f'}}\Delta_{f'}))}{\exp(\beta(x_{f'}+\frac{k}{n_{f'}}\Delta_{f'}))+\exp(\beta(x_f+\frac{j}{n_f}\Delta_f))} \end{align}
and
\begin{align}
& q_{f'_k,f_j}=\frac{\eta_{f_j}}{\exp(\tau)}\cdot
\frac{\exp(\beta(x_f+\frac{j}{n_f}\Delta_f))}{\exp(\beta(x_f+\frac{j}{n_f}\Delta_f))+\exp(\beta(x_{f'}+\frac{k}{n_{f'}}\Delta_{f'}))},
\end{align}
where $\sum_{j=-n_{f}}^{n_{f}}\eta_{f_j}=1$ and
$\sum_{k=-n_{f'}}^{n_{f'}}\eta_{f'_k}=1$.

Now we compute the stationary distribution of states for the
extended Markov chain $M'$. By detailed balance equation, we have
\begin{align}
p_{f_j}q_{f_j,f'_k}=p_{f'_k}q_{f'_k,f_j}, \forall j \in
\{-n_f,\ldots,n_f\}, k \in \{-n_{f'},\ldots,n_{f'}\}.
\end{align}

Then we have
\begin{align}
& p_{f_j}\cdot \frac{1}{\eta_{f_j} \cdot
\exp(\beta(x_f+\frac{j}{n_f}\Delta_f))} = p_{f'_k}\cdot
\frac{1}{\eta_{f'_k} \cdot \exp(\beta
(x_{f'}+\frac{k}{n_{f'}}\Delta_{f'}))}, \\ & \forall j \in
\{-n_f,\ldots,n_f\}, k \in \{-n_{f'},\ldots,n_{f'}\}. \nonumber
\end{align}

Therefore,
\begin{align}
\label{ratio_mc} \frac{p_{f_0}}{\eta_{f_0} \cdot \exp(\beta
x_f)}=\frac{p_{f'_0}}{\eta_{f'_0} \cdot \exp(\beta x_{f'})}
\end{align}
and
\begin{align}
\label{ratio_error}
\frac{p_{f'_k}}{p_{f'_0}}=\frac{\eta_{f'_k}}{\eta_{f'_0}} \cdot
\exp(\beta \frac{k}{n_{f'}} \Delta_{f'}), \forall k \in
\{-n_{f'},\ldots,n_{f'}\}.
\end{align}

Consider an arbitrary state $\hat{f}_0$ in the extended Markov chain
$M'$, where $\hat{f} \in \mathcal{F}$ and $\hat{f}\neq f,f'$. Since
state space of $M'$ is connected, we can always find a path to
connect $\hat{f}_0$ and $f_0$ through a series of adjacent states
$\tilde{f}(1)_0,\ldots,\tilde{f}(L)_0$, and $f_0=\tilde{f}(1)_0,
\tilde{f}(L)_0=\hat{f}_0$. Therefore,
\begin{align}
\frac{p_{\hat{f}_0}}{p_{f_0}}=\prod_{l=1}^{L-1}\frac{p_{\tilde{f}(l+1)_0}}{p_{\tilde{f}(l)_0}}
\end{align}

and by \eqref{ratio_mc} we have
\begin{align}
\frac{p_{\tilde{f}(l+1)_0}}{\eta_{\tilde{f}(l+1)_0} \cdot \exp(\beta
x_{\tilde{f}(l+1)})}=\frac{p_{\tilde{f}(l)_0}}{\eta_{\tilde{f}(l)_0}
\cdot \exp(\beta x_{\tilde{f}(l)})}.
\end{align}

Then
\begin{align}
\label{ratio_mc2} \frac{p_{\hat{f}_0}}{\eta_{\hat{f}_0} \cdot
\exp(\beta x_{\hat{f}})}=\frac{p_{f_0}}{\eta_{f_0} \cdot \exp(\beta
x_f)}.
\end{align}

By \eqref{ratio_error} and \eqref{ratio_mc2}, we know that $ \forall
f \in \mathcal{F}$,
\begin{align}
\label{db1} \frac{p_{f_0}}{\eta_{f_0} \cdot \exp(\beta x_f)} \text{
 is a constant}
\end{align}
and
\begin{align}
\label{db2} \frac{p_{f_j}}{p_{f_0}}=\frac{\eta_{f_j}}{\eta_{f_0}}
\cdot \exp(\beta \frac{j}{n_{f}} \Delta_{f}), \forall j \in
\{-n_{f},\ldots,n_{f}\}.
\end{align}

On the other hand, we have
\begin{align}
\label{db3} \sum\limits_{f \in \mathcal{F}} {\sum\limits_{j = - n_f
}^{n_f} {p_{f_j} } }  = 1.
\end{align}

By \eqref{db1}, \eqref{db2} and \eqref{db3}, we obtain the
stationary distribution of states for the extended Markov chain $M'$
as follows:
\begin{align}
& \forall f \in \mathcal{F}, j \in \{-n_{f},\ldots,n_{f}\},\nonumber\\
&\tilde{p}_{f_j}=\frac{\eta_{f_j} \cdot
\exp(\beta(x_f+\frac{j}{n_f}\Delta_f))}{\sum\limits_{f' \in
\mathcal{F}} \sum\limits_{k = - n_{f'} }^{n_{f'}} \eta_{f'_k} \cdot
\exp(\beta(x_{f'}+\frac{k}{n_{f'}}\Delta_{f'}))}.
\end{align}

The stationary distribution of peer configurations in the extended
Markov chain $M'$ is the probability distribution of aggregate
states $f_j, j \in \{-n_{f},\ldots,n_{f}\}$, i.e.,
\begin{align}
\bar{p}_f=\sum\limits_{j = - n_{f} }^{n_{f}} \tilde{p}_{f_j}.
\end{align}

Let \begin{align} \label{aver_coeff} \alpha_f \triangleq
\sum\limits_{j = - n_{f} }^{n_{f}}\eta_{f_j} \cdot
\exp(\beta\frac{j}{n_{f}}\Delta_{f}), \forall f \in \mathcal{F}.
\end{align}

Then we have
\begin{align}
\bar{p}_f=\frac{\alpha_{f}\exp(\beta x_{f})}{\sum\limits_{f' \in
\mathcal{F}} \alpha_{f'}\exp(\beta x_{f'})}, \forall f \in
\mathcal{F}.
\end{align}

By \eqref{dist_no_error}, we know
\begin{align}
p^*_f=\frac{\exp(\beta x_{f})}{\sum\limits_{f' \in \mathcal{F}}
\exp(\beta x_{f'})}, \forall f \in \mathcal{F}.
\end{align}

Let
\begin{align}
\label{aver_coeff2} \bar{\alpha}\triangleq \frac{\sum\limits_{f' \in
\mathcal{F}} \alpha_{f'}\exp(\beta x_{f'})}{\sum\limits_{f' \in
\mathcal{F}} \exp(\beta x_{f'})}.
\end{align}

It is not hard to see that $\frac{p^*_f}{\bar{p}_f}=
\frac{\bar{\alpha}}{\alpha_f}$, so
\begin{align}\label{bigdt}
p^*_f \ge \bar{p}_f \text{  iff  } \alpha_f \le \bar{\alpha}.
\end{align}

The total variation distance
\begin{align}
d_{TV}(\boldsymbol{p^*},
\boldsymbol{\bar{p}})&=\frac{1}{2}\sum\limits_{f \in
\mathcal{F}}|p^*_f -\bar{p}_f|\\ &= \sum\limits_{f \in A} (p^*_f
-\bar{p}_f),
\end{align}
where $A\triangleq \{f \in \mathcal{F}: p^*_f \ge \bar{p}_f \}$.

By \eqref{bigdt}, we know $A=\{f \in \mathcal{F}: \alpha_f \le
\bar{\alpha} \} \subset \mathcal{F}$.

Therefore, $\forall f \in A$,
\begin{align}
p^*_f - \bar{p}_f & = \frac{\exp(\beta x_{f})}{\sum\limits_{f' \in
\mathcal{F}} \exp(\beta x_{f'})} - \frac{\alpha_{f}\exp(\beta
x_{f})}{\sum\limits_{f' \in \mathcal{F}} \alpha_{f'}\exp(\beta
x_{f'})} \\ & = \frac{\exp(\beta x_{f})}{\sum\limits_{f' \in
\mathcal{F}} \exp(\beta x_{f'})} - \frac{\alpha_{f}\exp(\beta
x_{f})}{\bar{\alpha}\sum\limits_{f' \in \mathcal{F}} \exp(\beta
x_{f'})} \\ & = \frac{\exp(\beta x_{f})}{\sum\limits_{f' \in
\mathcal{F}} \exp(\beta x_{f'})} [1-\frac{\alpha_{f}}{\bar{\alpha}}].
\label{difference}
\end{align}

Since $\sum\limits_{j = - n_{f} }^{n_{f}} \eta_{f_j}=1$ and $\forall
j \in \{-n_{f},\ldots,n_{f}\}$,
\begin{align}
& \exp(\beta\frac{j}{n_{f}}\Delta_{f}) \ge \exp(-\beta\Delta_{f})
\ge \exp(-\beta\Delta_{\max})
\end{align}
and
\begin{align}
& \exp(\beta\frac{j}{n_{f}}\Delta_{f}) \le \exp(\beta\Delta_{f}) \le
\exp(\beta\Delta_{\max}),
\end{align}
by \eqref{aver_coeff} we know that $\forall f \in \mathcal{F}$
\begin{align}
& \alpha_f \ge \sum\limits_{j = - n_{f} }^{n_{f}} \eta_{f_j} \cdot
\exp(-\beta\Delta_{\max}) = \exp(-\beta\Delta_{\max})
\end{align}
and
\begin{align}
& \alpha_f \le \sum\limits_{j = - n_{f} }^{n_{f}} \eta_{f_j} \cdot
\exp(\beta\Delta_{\max}) = \exp(\beta\Delta_{\max}).
\end{align}

Then by \eqref{aver_coeff2}, we have $\bar{\alpha} \le \exp(\beta
\Delta_{\max})$. Therefore,
\begin{align}
1-\frac{\alpha_{f}}{\bar{\alpha}} \le 1-\frac{\exp(- \beta
\Delta_{\max})}{\exp(\beta\Delta_{\max})} = 1- \exp(-2 \beta
\Delta_{\max}) , \forall f \in A \subset \mathcal{F}.
\end{align}

So by \eqref{difference}, we have $\forall f \in A$,
\begin{align}
p^*_f - \bar{p}_f & = \frac{\exp(\beta x_{f})}{\sum\limits_{f' \in
\mathcal{F}} \exp(\beta x_{f'})} [1-\frac{\alpha_{f}}{\bar{\alpha}}]
\\ & \le \frac{\exp(\beta x_{f})}{\sum\limits_{f' \in
\mathcal{F}} \exp(\beta x_{f'})} (1-\exp(-2 \beta \Delta_{\max})).
\end{align}

Then,
\begin{align}
d_{TV}(\boldsymbol{p^*}, \boldsymbol{\bar{p}})& = \sum\limits_{f \in
A} (p^*_f -\bar{p}_f) \\ & \le \sum\limits_{f \in A}
\frac{\exp(\beta x_{f})}{\sum\limits_{f' \in \mathcal{F}} \exp(\beta
x_{f'})} (1-\exp(-2 \beta \Delta_{\max})) \\ & \le \sum\limits_{f
\in \mathcal{F}} \frac{\exp(\beta x_{f})}{\sum\limits_{f' \in
\mathcal{F}} \exp(\beta x_{f'})} (1-\exp(-2 \beta \Delta_{\max})) \\
& = 1-\exp(-2 \beta \Delta_{\max}).
\end{align}

Therefore,
\begin{align}
|\boldsymbol{p^*} \boldsymbol{x}^T - \boldsymbol{\bar{p}}
\boldsymbol{x}^T| & = |\sum\limits_{f \in \mathcal{F}} (p^*_f
-\bar{p}_f)x_f | \\ & \le x_{\max} \sum\limits_{f \in \mathcal{F}}
|(p^*_f -\bar{p}_f)| \\ & = 2x_{\max}d_{TV}(\boldsymbol{p^*},
\boldsymbol{\bar{p}}) \\ & \le 2x_{\max}(1-\exp(-2 \beta
\Delta_{\max})).
\end{align}

This concludes the proof.

\end{document}